\documentclass[%
superscriptaddress,
pre,
 amsmath,amssymb,
 aps,
twocolumn     
]{revtex4-2}

\usepackage{graphicx}
\usepackage{dcolumn}
\usepackage{bm}
\usepackage[usenames]{color} 

\newcommand{\be}{\begin{equation}}
\newcommand{\ee}{\end{equation}}
\newcommand{\bea}{\begin{eqnarray}}
\newcommand{\eea}{\end{eqnarray}}

\newcommand{\Eb}{E_\text{b}}
\newcommand{\Esub}{E_\text{sub}}
\newcommand{\DE}{\Delta E}
\newcommand{\DEm}{{\Delta E}_\text{m}}
\newcommand{\DEc}{{\Delta E}_\text{c}}
\newcommand{\nuc}{\nu_\text{c}}

\newcommand{\nus}{\nu_\text{sub}}
\newcommand{\Rs}{R_\text{sub}}
\newcommand{\ks}{k_\text{sub}}

\newcommand{\Nisl}{n_\text{isl}}
\newcommand{\Nm}{N^{(1)}}
\newcommand{\Nc}{N_\text{c}}
\newcommand{\Estar}{E^*}

\newcommand{\OP}{\Psi_\text{1-2}}

\begin{document}


\title{Island formation in heteroepitaxial growth}

\author{Frederik Munko}
\affiliation{Max Planck Institut f{\"u}r Polymerforschung, Ackermannweg 10, 55128 Mainz, Germany}
\affiliation{Eberhard Karls Universit{\"at} T{\"u}bingen, Institut f{\"u}r Angewandte Physik, Auf der Morgenstelle 10, 72076 T{\"u}bingen, Germany }
\author{Catherine Cruz Luukkonen}
\affiliation{Eberhard Karls Universit{\"at} T{\"u}bingen, Institut f{\"u}r Angewandte Physik, Auf der Morgenstelle 10, 72076 T{\"u}bingen, Germany }
\author{Ismael S. S. Carrasco}
\affiliation{International Center of Physics, Institute of Physics, University of Brasilia, 70910-900 Brasilia, Federal District, Brazil}
\author{F\'abio D. A. Aar{\~{a}}o Reis}
\email{fdaar@protonmail.com}
\affiliation{Instituto de F\'{\i}sica, Universidade Federal Fluminense, Avenida Litor\^{a}nea s/n, 24210-340 Niter\'{o}i, RJ, Brazil}
\author{Martin Oettel}
\email{martin.oettel@uni-tuebingen.de}
\affiliation{Eberhard Karls Universit{\"at} T{\"u}bingen, Institut f{\"u}r Angewandte Physik, Auf der Morgenstelle 10, 72076 T{\"u}bingen, Germany }


\date{\today}

\begin{abstract}   

{Island formation in strain--free heteroepitaxial deposition of thin films is analyzed using kinetic Monte Carlo simulations of two minimal lattice models and scaling approaches.
The transition from layer-by-layer (LBL) to island (ISL) growth is driven by a weaker binding strength of the substrate which, in the kinetic model, is equivalent to an increased diffusivity of particles on the substrate compared to particles on the film.
The LBL-ISL transition region is characterized by particle fluxes between layers 1 and 2 significantly exceeding the net flux between them, which sets a quasi-equilibrium condition.
Deposition on top of monolayer islands weakly contributes to second layer nucleation, in contrast with the homoepitaxial growth case.
A thermodynamic approach for compact islands with one or two layers predicts the minimum size in which the second layer is stable.
When this is linked to scaling expressions for submonolayer island deposition, the dependence of the ISL--LBL transition point on the kinetic parameters qualitatively matches the simulation results, with quantitative agreement in some parameter ranges.
The transition occurs in the equilibrium regime of partial wetting and the convergence of the transition point upon reducing the deposition rate is very slow and practically unattainable in experiments.}       


\end{abstract} 

\maketitle   


\section{Introduction}

The integration of different nanostructured materials is essential for the fabrication of novel devices for electronics, energy storage, and clean energy production, among others \citep{nano_for_energy_2013,sunAdvElMat2019,liuAM2021,zhengSciAdv2021,yuanAdvElMat2021}.
These materials are frequently built as thin solid films whose morphologies must be tailored to fit the desired applications \cite{hinderhoferCPC2012,pereira_2021,leo_rev_2023}.
The required film thicknesses may decrease to a few nanometers, but the coverage requirements of comparably much larger lengths, rapid growth, and temperature control remain.
For this reason, the knowledge of the nonequilibrium growth conditions that lead to initial layer--by--layer (LBL) growth (Frank--van der Merwe mode) or to island (ISL) formation (Vollmer--Weber mode) is essential \citep{ohring}.
In the last two decades, this knowledge acquired increasing importance for the improvement of thermal vapor deposition of organic films on amorphous substrates such as $\text{SiO}_x$ \citep{kowarikJPCM2008,lorchAPL2015,dullMatHor2022}.
For instance, ISL growth is observed in films of $\text{C}_{60}$ \cite{Singh_2007_ApplPhysLett,Yim_2009_ApplPhysLett,Reisz_2021}, but LBL followed by ISL growth (termed Stranski--Krastanov mode \citep{ohring}) occurs in pentacene (PEN) \cite{kowarik2007} and diindenoperylene (DIP) \cite{kowarikPRL2006,frankPRB2014a} films.

ISL formation is observed both in homoepitaxial and heteroepitaxial film growth, but its origin may be very different.
In homoepitaxial growth the substrate is of the same material as the growing film and usually there is an initial layer--by--layer (LBL) or Frank--van der Merwe growth, followed by roughening or mound formation \cite{venables,michely,etb}.
This type of roughening is often related to the so--called Ehrlich--Schw{\"obel} (ES) barrier, which is a potential barrier for step--edge (interlayer) diffusion \cite{ehrlich,ES}.
In heteroepitaxial growth, the substrate may differ from the growing material both in terms of the underlying lattice parameters (a mismatch if substrate and film are crystalline) and in the binding energies (the one--body potential of a flat substrate and a flat film exerted on a film atom/molecule are different).
In the former case, the lattice parameters of the growing film may initially fit those of the substrate, but internal or residual stresses are released when thicker films adopt their equilibrium structures \citep{pimpinelli}.
However, this mechanism is expected to be absent if the substrate is amorphous and may be negligible in other systems with weak substrate/film interactions \citep{kowarikJPCM2008,lorchAPL2015,dullMatHor2022,elofsson2014}.
Thus, the different binding energies of substrate/film and film/film particles must play the main role.

In this context, ISL formation is very often pictured as a quasi--equilibrium process where islands are droplets of film material on a partially wetting substrate \citep{venables,burkeJPCM2009}.
For partial wetting, the surface tensions between substrate and vacuum/gas (above the film) and between flat film and vacuum/gas are unequal due to the different binding energies.
In this quasi--equilibrium picture, the transition from ISL to LBL growth would correspond to the equilibrium wetting transition for which temperature (experimentally easily accessible) and the binding energy difference (theoretically well suited) are control parameters.
Additional complications may arise in the extension to the nonequilibrium conditions of heteroepitaxial deposition.

Plentiful theoretical insight into homoepitaxial thin film deposition has has been gained through scaling considerations and rate equation approaches \citep{pimpinelli,michely,etb,misbah2010}, many of them checked and validated by kinetic Monte Carlo (KMC) simulations of particle--based models with restricted sets of particle moves between lattice sites \citep{cv,etb}.
Previous work on ISL formation was concentrated on the second layer nucleation on top of the first layer, in which one of the most important elements is the ES reduced step--edge motion downwards \citep{tersoffPRL1994,krugPRB2000,heinrichsPRB2000,krugEPJB2000}.
The interplay with the necessarily corresponding step--edge moves upwards did not play a major role and in many cases was simply neglected in the modeling.
Instead, heteroepitaxial growth changes the substrate/film particle binding energies, which will directly influence the step--edge moves upwards from the substrate to the film and which may be a major driving force for ISL formation {(independently of ES barriers)}.
The importance of this mechanism was recently demonstrated in KMC simulations that qualitatively describe the initial stages of deposition of CdTe films \citep{toApplSS2021,toreis2022}, of several organic films \citep{empting2021,empting2022}, and of silver films \citep{luPRMat2018,gervilla2019} on a variety of substrates.
However, a conceptual explanation of the LBL--ISL transition is still lacking.

This paper addresses that transition in the initial stages of heteroepitaxial growth by a combination of three approaches:
(i) KMC simulations of two atomistic models of thin film deposition with negligible ES barriers;
(ii) an equilibrium model that accounts for the dominant role of hops from the substrate to the film surface (which is supported by the KMC data);
(iii) previous numerical and analytical results on nonequilibrium submonolayer growth \citep{ratsch1995,barteltSSL1995,amarSS1997,oliveirareis2013}.
The atomistic models have been recently introduced and some phenomenology of ISL growth has been established \cite{toApplSS2021,empting2021}.
Both models have essentially one additional parameter to characterize the heteroepitaxy when compared to their homoepitaxy counterparts, namely a diffusion coefficient on the substrate \citep{toApplSS2021} or a substrate/film energy \citep{empting2021}.
With an appropriate connection between the parameters of these models, they provide similar descriptions of the LBL--ISL transition with transition points (``dynamic wetting'' points) that may be far from those of equilibrium wetting.
The KMC simulations and scaling arguments for compact island formation show that second--layer atoms mainly originate from step moves from the substrate and not from deposition.
This allows us to assume local equilibration of the height of single islands during submonolayer growth in the theoretical approach that predicts when the LBL--ISL transition takes place before those islands merge.
The approach is consistent with the simulation data and shows that the equilibrium wetting limit is reached very slowly with increasing diffusion-to-deposition ratio, rationalizing the findings on the dynamic wetting transition.

\section{Models and methods}
\label{modelmethods}

\subsection{Two minimal models for heteroepitaxy}
\label{modelsection}

The possible particle positions are sites of a simple cubic lattice and the unit of length is given by the edge of the cubic unit cell.
The substrate is the plane $z=0$ and solid--on--solid (SOS) conditions (no overhangs) are imposed on the film.
Particles are deposited at the top of randomly chosen columns $\left( x,y\right)$ with a rate $F$ per lattice site.
For simplicity, we set $F=1$ in the simulations, so that the time unit is the deposition time of $1$ monolayer (ML).

Particles at $z=1$ interact with the substrate of a different material and with particles that occupy lateral nearest neighbor (NN) sites, whereas particles at $z\geq 2$ interact with the same species in all occupied NN sites.
Particle hops are performed to neighboring columns in $\pm x$ or $\pm y$ directions.
All energies involved in the hopping rates are given in units of the thermal energy $k_\text{B}T$.\newline

\textit{Clarke--Vvedensky (CV) model \citep{cv}. --} A base hopping rate $k$ applies to particles with $z\geq2$ that have no lateral nearest neighbors (NNs), which are those in a plane with film particles below.
Free diffusion on the substrate ($z=1$) is characterized by a base rate $\ks$.
Associating an attractive energy $-\Eb$ to lateral bonds (same $z$), the hopping rate receives an extra factor of $\exp(-\Eb)$ for each lateral bond the particle has in its initial state.
The base rates $k$ and $\ks$ may be different and associated with activation energies $E_\text{D}$ and $\Esub$, respectively, and the same attempt frequency $f$ \citep{toApplSS2021}.
Physically, this is linked to different diffusion barriers for a particle to hop on different materials, namely the substrate and the film.
ES barriers are neglected in this minimal version of the CV model.

\begin{figure}[!h]
\includegraphics[width=0.48\textwidth]{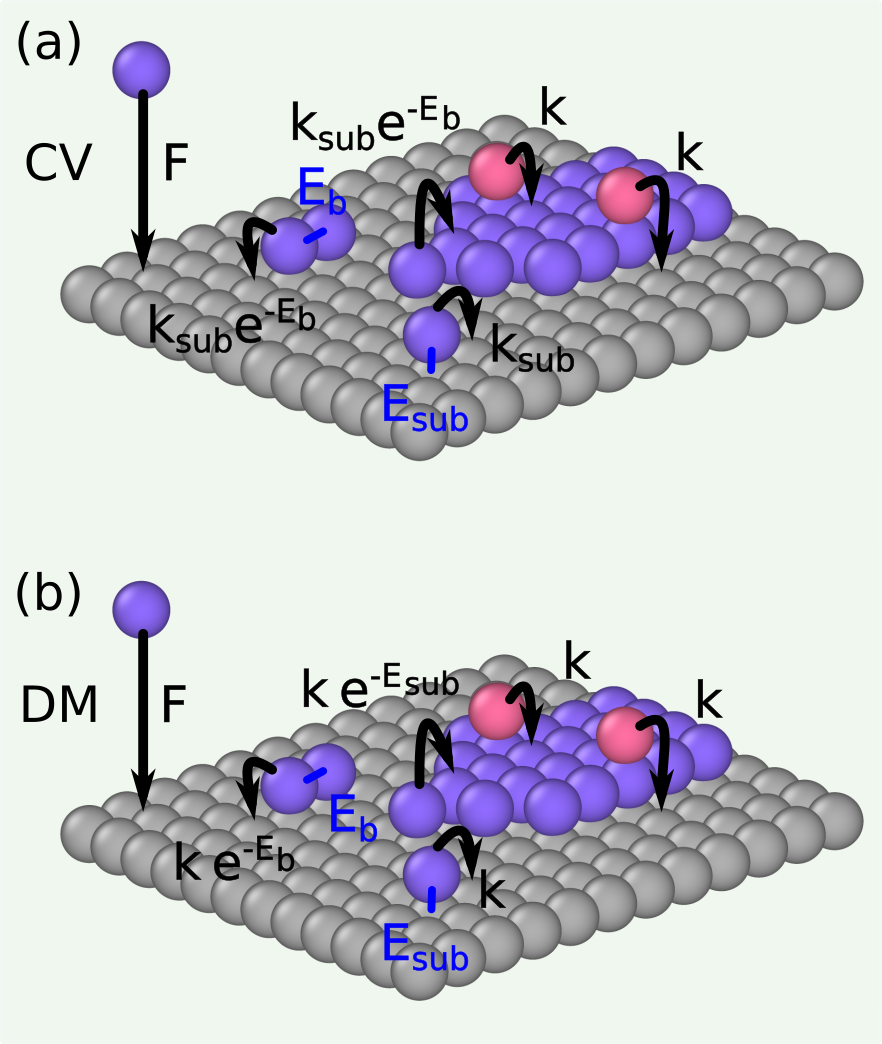}
  \caption{Basic hops and their corresponding rates in the (a) CV and (b) DM models. }
\label{fig:hops}
\end{figure}

In this work we set $E_\text{D}=\Eb$, i.e. equal energies for vertical and lateral bonds between deposited particles.
This is suitable for connections with the DM model (defined below) and with a theoretical approach to the LBL-ISL transition (Sec. \ref{sec:dynwet}) and leads to
\be
\begin{split}
k & =f \exp{\left(-\Eb\right)} \quad , \\
\ks & =f \exp{\left(-\Esub\right)} = \nu k \quad ,
\end{split}
\label{kCV}
\ee
where
\be
\nu= \exp{\left(\DE\right)} \quad , \quad  \DE=\Eb - \Esub .
\label{ECV}
\ee
Here $\nu$ is a ratio of particle diffusivities on the substrate ($z=1$) and on the film ($z\geq2$); $\nu>1$
is expected for particle-substrate interactions weaker than those between two particles, i.e. $\DE>0$.
$\nu=1$ ($\DE=0$) is characteristic of homoepitaxial growth.
Fig.~\ref{fig:hops}(a) illustrates the basic moves and their rates in this model. 

Free diffusion of particles with a base rate $k$ is characterized by a mean square displacement in the plane $\langle r^2 \rangle = 4 kt$, so that the diffusion constant is $D=k$ in units of the lattice unit area per unit time.
We denote the characteristic diffusion-to-deposition ratios $R$ on the film ($z\geq2$) and $\Rs$ on the substrate ($z=1$) as
\be
R=\frac{k}{F} \quad , \quad \Rs =\frac{\ks}{F} =  \nu R .
\label{defR}
\ee

\textit{Dynamic Metropolis (DM) model \citep{empting2021}. --} 
Neighboring particles (laterally or vertically) are bonded with an attractive energy $-\Eb$ whereas particles on the substrate are bonded to that substrate with an attractive energy $-\Esub$.
The film relaxation is implemented via a Metropolis--like energetic algorithm.
The hops to neighboring lateral sites have a rate $k\, \text{min} (1, \exp(-\delta E))$, where $k$ is a base rate and $\delta E$ is the energy difference between the final state (after the hop) and the initial state (before the hop).
This base rate $k$ applies to freely diffusing particles (no lateral bonds) on a plane at any $z>0$, i.e. either substrate or film.
In their lateral moves, particles can move up at most one layer (if the neighboring site is not blocked) or move down at most one layer.
See Fig.~\ref{fig:hops}(b) for an illustration of the basic moves and their rates.
ES barriers are also neglected in this model.

The energetics of the DM model is equivalent to that of the CV model when the condition $E_\text{D}=\Eb$ is assumed.
The relaxation to equilibrium by particle diffusion obeys detailed balance in both models and the equilibrium configurations are the same.
However, the diffusion-to-deposition ratio $R$ applies to particles in all layers in the DM model, while the difference in their interactions with other particles and with the substrate is contained in the energy difference $\DE$.
Consequently, the relaxation to equilibrium provided by the CV and by the DM model is different.
The random deposition is the mechanism that continuously deviates the growth from that global equilibration.

\subsection{Quantities of interest and simulation details}
\label{simulations}

A unit flux $F$ is set, so the number of deposited monolayers is numerically equal to the deposition time $t$.
The coverage of layer $i$ at time $t$, $\Psi_i(t)$, is defined as the fraction between the number of particles in that layer and the total number of sites of that layer.

For characterizing the LBL--ISL transition, an order parameter $\OP$ was recently defined as \citep{empting2021}
\be
  \OP = \Psi_1(t=1) - \Psi_2(t=1)\;,
\ee 
which is the difference in coverages of the first and second layer after deposition of one monolayer of material ($t=1$).
For ideal LBL growth, the order parameter is $1$, i.e. the substrate is fully covered and there is no particle in the second layer.
For ideal ISL growth, $\OP=0$ since islands of minimal energy have ``vertical'' walls with an equal number of particles in the first and second layers (this does not exclude the possibility of occupation of higher layers).

The LBL--ISL transition can also be characterized by the particle flux between different layers.
The flux $\Phi_{i\to j}$ from layer $z=i$ to layer $z=j$ is defined as the number of particle hops per substrate site per unit time.
We are particularly interested in the particle exchange between layer $1$ and the upper layers, so we denote the net flux to layer $1$ as
\be
\Phi_{\text{net},1}=\sum_{i\geq2}{\Phi_{i\to 1} - \Phi_{1\to i}} .
\label{Phinet}
\ee
Positive $\Phi_{\text{net},1}$ indicates dominant flux downwards and is typical of LBL growth, whereas negative $\Phi_{\text{net},1}$ indicates particle accumulation in the upper layers and a trend to ISL formation.
The comparison of $\Phi_{1\to i}$ and $\Phi_{i\to 1}$ for each $i\geq2$ is also useful for the interpretation of the process of island formation.

In ideal LBL growth, the net flux $\Phi_{\text{net},1}$ has a particularly simple time evolution.
Only the layers $z=1$ and $z=2$ are relevant in this case.
At time $t$, the first layer coverage is $\approx t$, so the number of particles per site deposited on $z=2$ in a time interval $\Delta t$ is ${\Delta N}_{\text{dep},2}=t \Delta t$
(within $\Delta t$, the total increase in coverage  [particles per site]  is $\Delta t$, and
the fraction $t$ of it goes into layer 2).
All these particles must go downwards in this time interval (otherwise a second layer would be formed).
Moreover, all particles that hop from $z=1$ to $z=2$ also have to hop down in a short time interval.
Thus, the net flux is that of particles deposited on the second layer:
\be
\Phi_{\text{net},1}={\Delta N}_{\text{dep},2}/\Delta t \approx t \quad \text{(ideal LBL)} .
\label{LBLflux}
\ee

The simulations of the CV and DM models were performed with $R$ varying from ${10}^3$ to ${10}^7$ and with
$\nu=\Rs/R$ ranging from $1$ to $\sim{10}^2$ depending on the choice of the other parameters.
The interaction energy between NN particles range between $\Eb=3$ and $9$.
For each parameter set, averages were performed among $10$ realizations (for Fig.~\ref{fig:comparison}), otherwise 5.
The deposition time was restricted to $t\leq 2$ in most simulations.
The lateral size of the simulation cell was $L=256$.
The simulations were implemented with the kinetic Monte Carlo (kMC) algorithm described in Ref. \citep{carrascoPRE2023}.


\section{Fundamentals of previous works}
\label{fundamentals}

\subsection{Equilibrium wetting transition}
\label{equilibriumtransition}

Here we recall the thermal equilibrium properties of the model without particle deposition, in which particle-particle interactions have energy $-\Eb$ and particle-substrate interactions have energy $-\Esub$.
This lattice gas model is equivalent to an Ising model at conditions allowing for wetting and layering transitions \citep{binder1992}, as recently reviewed in Ref. \protect\cite{empting2021} in the context of growth.

At $\Eb\approx1.64$, there is a roughening transition because the particle-particle interaction is so small (relative to the thermal energy) that steps are created on the film with no free energy cost.
However, thin film deposition must be performed at temperatures below that transition, so we can generally consider $\Eb\geq2$.
In these conditions, an equilibrium wetting transition takes place approximately at $\Esub/\Eb=1$ and can be interpreted in purely energetic terms (equivalent to zero temperature conditions).
For $\Esub\geq\Eb$, the particle-substrate interaction is high and the film wets the substrate.
This is equivalent to $\nu\leq1$ in our {kinetic models CV and DM} [Eq. (\ref{ECV})];
however, only the case $\nu=1$ of this equilibrium wetting phase is considered in the present work.
For $\Esub$ only slightly smaller than $\Eb$, the weaker particle-substrate interaction produces partially-wetted films, i.e. the film breaks into islands that cover the substrate only in parts.
Thus, in the conditions $\nu>1$ considered here for heteroepitaxy modeling, the partially wetted phase is the equilibrium configuration.

\subsection{Scaling regimes of submonolayer growth}
\label{submonolayer}

Here, we review previous results from rate equation theories and simulations of submonolayer growth that aid in interpreting the initial stages of film deposition.
This review is limited to the CV model.

The studies of submonolayer growth with the CV model considered particle occupation restricted to the first layer and coverages $\lesssim30\%$, which are usually below island coalescence \citep{venables1987,ratsch1995,barteltSSL1995,amarSS1997,oliveirareis2013}.
The scaling regimes of the island density and free particle density match those of rate equation theories that set critical sizes above which the islands are stable \citep{venables1987,barteltSSL1995}; see also Ref. \protect\cite{etb}.
In the $\Rs$–$\Eb$ parameter plane (as shown later in Fig.~\ref{fig:Rsub_Eb}), two scaling variables distinguish the aggregation regimes in square lattices (substrates of simple cubic lattices):
\be 
\begin{split}
  Y_1 & = \exp(-\Eb)\,\Rs^{2/3} \quad , \\
  Y_2 & = \exp(-\Eb)\,\Rs^{1/5} \quad.
\end{split}
  \label{defY}
\ee 
The observed regimes are:

\par\noindent (I) Stable dimers, in which particles with a single lateral bond are immobile during the monolayer growth, which appear for
\be 
  Y_1 \lesssim 1  \quad \leftrightarrow \quad \Eb \gtrsim \frac{2}{3} \ln{\Rs} \;.
\ee 
The formation of highly branched islands resembling diffusion-limited-aggregates (DLAs) \citep{witten} is observed when $Y_1\ll1$ and the crossover to regime II below occurs in a wide region $1\lesssim Y_1\lesssim10$.

\par\noindent (II) Stable squares ($2\times2$ islands), in which particles with one lateral bond can easily hop but particles with $2$ or more lateral neighbors are frozen while the monolayer grows.
The formation of compact islands with wide flat borders is typical in this interval.
This regime is fully developed if
\be
\begin{split}
  Y_1\; & \gtrsim\;10 \quad\text{and} \quad Y_2 \;\lesssim\; 0.3 \quad \leftrightarrow \\
  \Eb\; & \lesssim\; -\ln(10) +\frac{2}{3} \ln{\Rs}\; \quad\text{and} \\
  \Eb\; & \gtrsim\; -\ln(0.3) +\frac{1}{5} \ln{\Rs} \;.
\end{split}
\ee

\par\noindent (III) Diffuse islands, in which particles with $2$ lateral bonds likely hop in the timescale of one-layer deposition.
This regime appears for $Y_2\gtrsim1$.
Rapid particle exchange between neighboring islands and their coarsening are features of regime III \citep{oliveirareis2013}.

In the CV model, all bonds may be eventually broken and no island is actually stable.
However, the equivalence with rate equation theories is possible because the stability criteria of the CV model are based on the probabilities of breaking island bonds in the timescale of deposition of a monolayer.
This equivalence can be observed only if the hopping rate changes by a factor $\ll1$ when the number of lateral bonds increases by one unit, which typically requires $\Eb\gtrsim5$ (and the corresponding Boltzmann factor $\lesssim0.01$).

\begin{figure*}[t]
  \includegraphics[width=\textwidth]{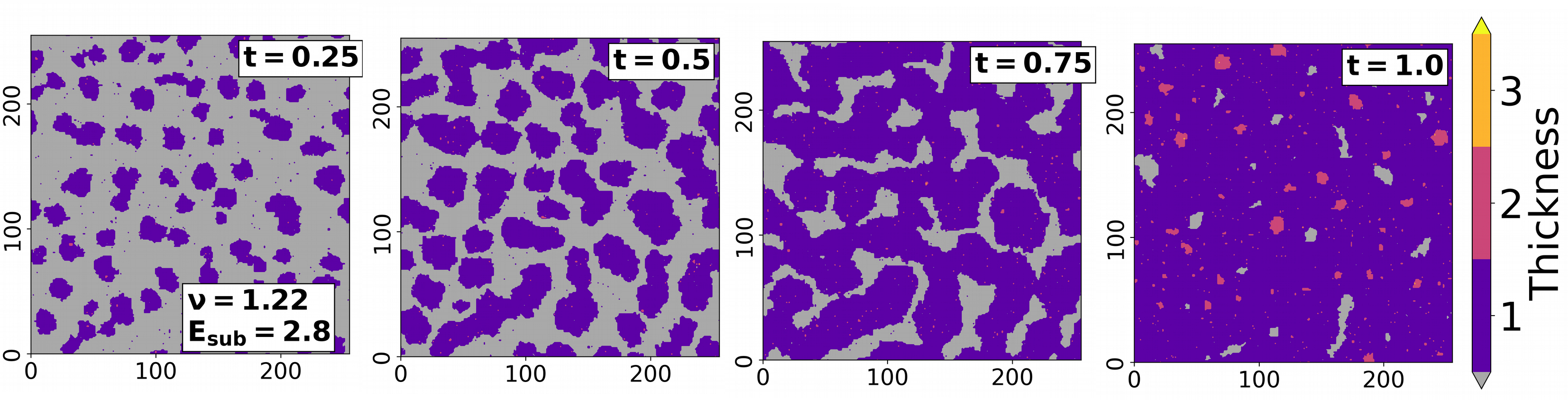}
  \caption{LBL growth: Sequence of snapshots of deposits grown with the CV model with $R=10^5$, $\Eb=3$, $\nu=1.22$ (intervals of $0.25$~ML between snapshots).
}
  \label{fig:snap_sequenceLBL}
\end{figure*}

\section{Numerical study of island formation}
\label{secondlayer}

\subsection{Morphology evolution}
\label{morphology}

Figures \ref{fig:snap_sequenceLBL} shows a sequence of snapshots from CV model simulations with $R=10^5$, $\Eb=3$, and $\nu=1.22$ ($\DE=0.20$), i.e. with a particle diffusion coefficient on the substrate only slightly larger than that on $z\geq 2$ (Eq.~(\ref{kCV})), equivalent to a slightly less attractive substrate as expressed by the low value of $\Delta E$.
The first layer islands grow and coalesce with a small filling of the second layer, so this is a typical case of LBL growth.
This differs from the partial wetting phase (ISL) obtained in thermal equilibrium with the same $\Eb$ and $\DE$.

\begin{figure*}[t]
  \includegraphics[width=\textwidth]{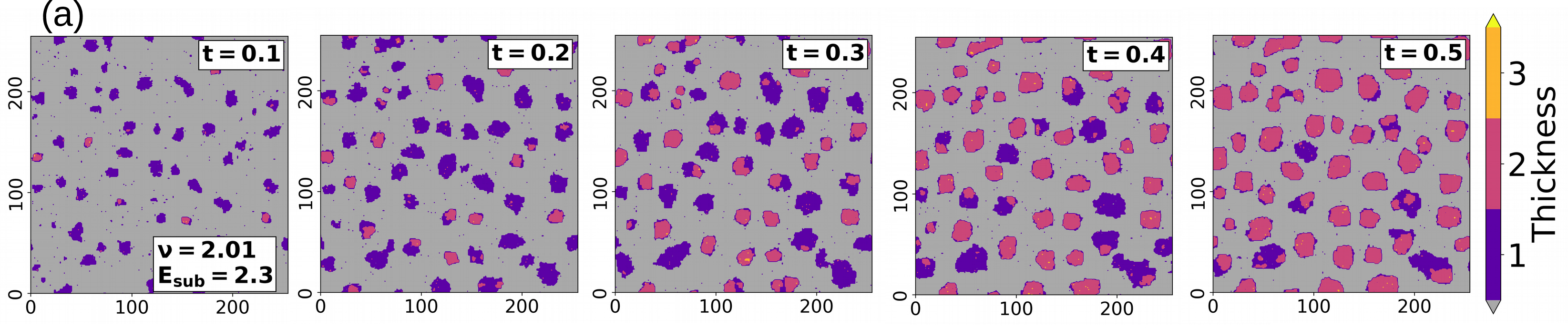}\\
  \includegraphics[width=\textwidth]{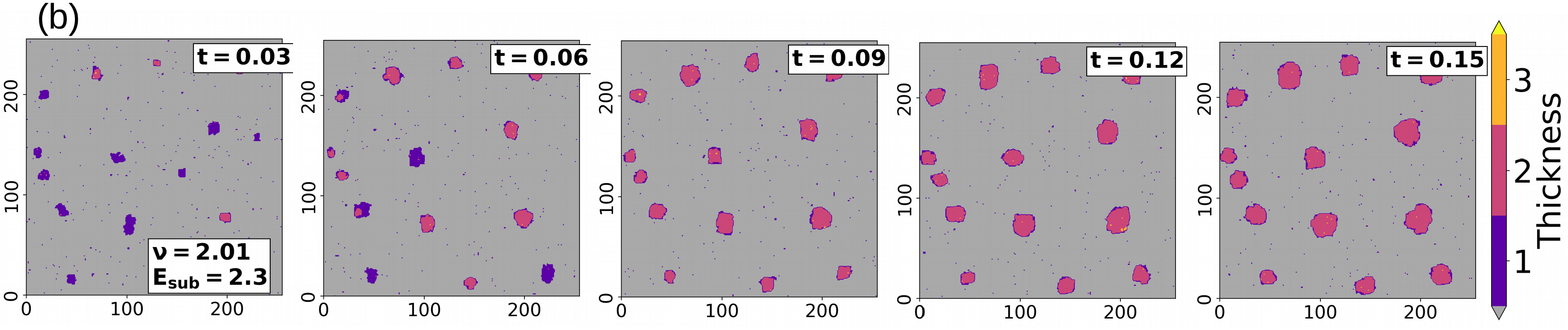}
  \includegraphics[width=\textwidth]{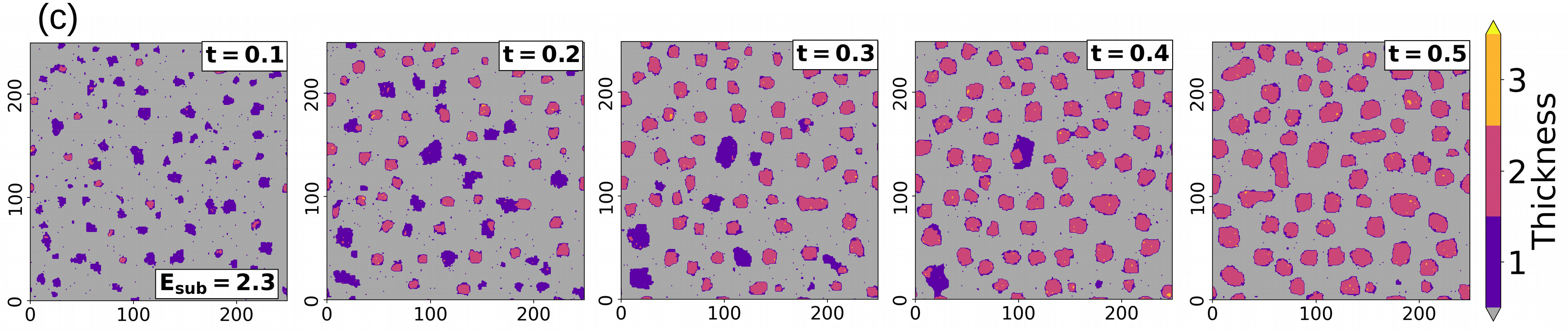}
  \caption{ISL growth: Sequence of snapshots of deposits grown with:
(a) CV model with $R=10^5$, $\Eb=3$, $\nu=2.01$ (intervals of $0.1$~ML);
(b) CV model with $R=10^6$, $\Eb=3$, $\nu=2.01$ (intervals of $0.03$~ML);
(c) DM model with $R=10^5$, $\Eb=3$, $\exp(\Delta E)=2.01$ (corresponding to $\nu$) (intervals of $0.1$~ML).
}
  \label{fig:snap_sequence}
\end{figure*}

Figure \ref{fig:snap_sequence}(a) shows snapshots from simulations with the same values of $R$ and $\Eb$ but with $\nu=2.01$ ($\DE=0.7$).
With this small increase in the particle diffusivity on the substrate, two--layer islands are formed in the early growth stages and second--layer formation has been almost completed at a deposition of $0.5$~ML (when tiny islands at the third layer are also present).
At this point, the uncovered first layer particles are mostly free.
This anticipates a sharp LBL--ISL transition as $\nu$ (or $\DE$) increases.

Figure \ref{fig:snap_sequence}(b) shows snapshots from CV model simulations with $R=10^6$, $\Eb=3$, and $\nu=2.01$ ($\DE=0.7$).
The increase of the diffusion-to-deposition ratio shortens the second layer completion to about $0.12$~ML.
Again, the uncovered first layer particles are mostly free particles, whose diffusion coefficient is $\sim20$ times larger than that of particles bound to a single NN in an island border (and a factor $\sim400$ in comparison with particles bound to two NNs).
This gives evidence that the individual process of formation of a two--layer island is (a) a rather fast process ending at a small, $R$--dependent submonolayer coverage and (b) a collective process which is presumably mainly driven by energy minimization to reach the compact two--layer shape.

Figure \ref{fig:snap_sequence}(c) shows a sequence of snapshots from DM model simulations with the same dimensionless parameters as in Figure \ref{fig:snap_sequence}(a).
Both sequences are quite similar, although the second layer formation is completed a bit faster in the DM model.
In DM model simulations with $\nu=1.22$, almost ideal LBL growth is observed, similarly to Fig.~\ref{fig:snap_sequenceLBL}.
Moreover, increasing $R$ to ${10}^6$ with $\Eb=3$ and $\nu=2.01$ leads to second layer completion at much shorter times compared to that of Fig.~\ref{fig:snap_sequence}(c), in parallel with our observations in CV model simulations.

Notably, the second layer nucleation shown here is controlled by the weakening of particle-substrate interactions without ES barriers, as anticipated in models of deposition of films of semiconductors, organic molecules, and metals \citep{toApplSS2021,toreis2022,empting2021,empting2022,luPRMat2018,gervilla2019}.
These conditions are very different from those of second--layer nucleation in homoepitaxial deposition which considered high ES barriers \citep{tersoffPRL1994,krugPRB2000,heinrichsPRB2000,krugEPJB2000}.

\subsection{LBL--ISL transition}

\begin{figure}[!h]
  \includegraphics[width=0.45\textwidth]{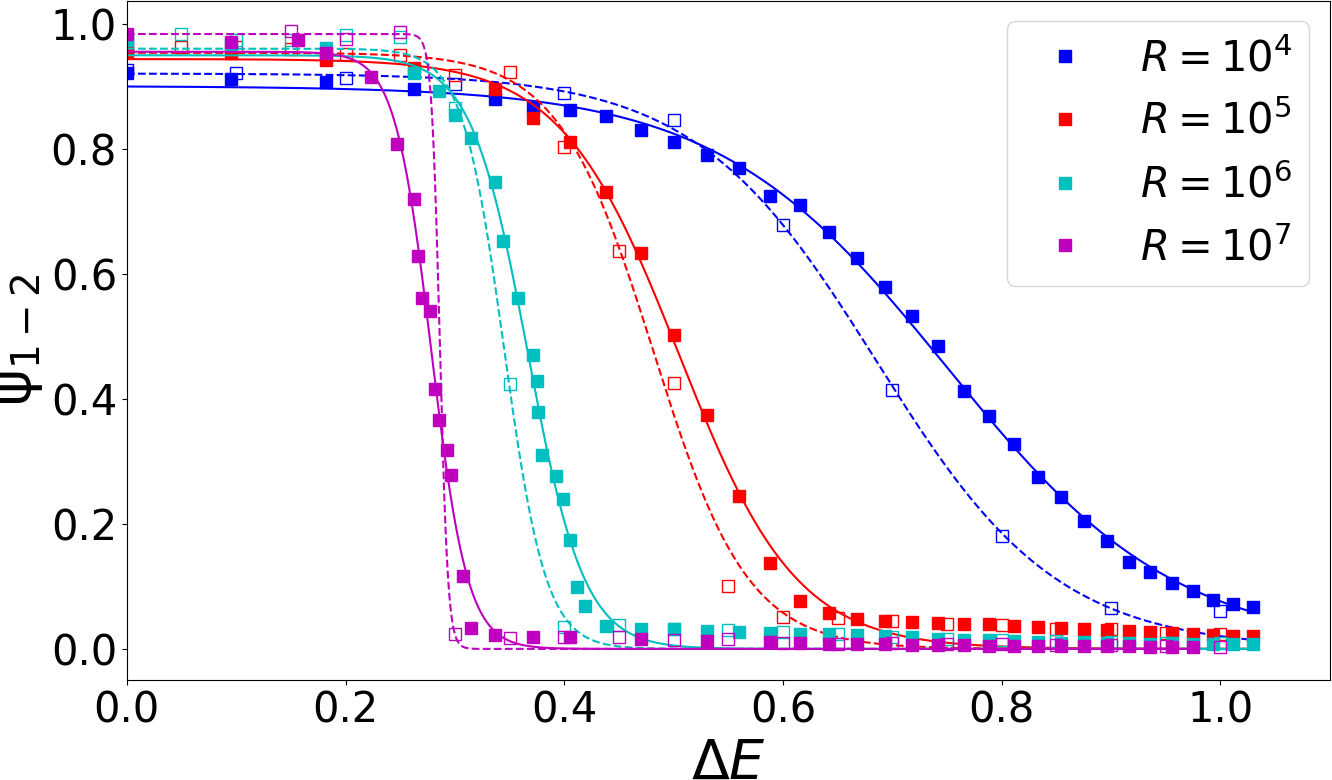}
  \caption{Comparison between the CV model (filled symbols, fits [Eq.~\ref{OPfit}] full lines) and DM model (open symbols, fits dashed lines) for the behavior of the order parameter $\OP$.
  }
  \label{fig:comparison}
\end{figure}

Fig.~\ref{fig:comparison} shows the order parameter curves $\OP(\DE)$ obtained in simulations of the CV model (filled symbols, fits by full lines [Eq.~(\ref{OPfit}) below]) and of the DM model (open symbols, fits by dashed lines) for various $R$ and constant binding energy $\Eb=3$.
Here the control variable $\DE$ is used instead of $\nu$ because the former allows simpler and more accurate fitting of the order parameter curves.

The case $\DE=0$ ($\nu=1$) is the model of homoepitaxial deposition with negligible ES barriers.
It leads to $0.9<\OP<1$, corresponding to LBL growth.
This happens because the mobility of first--layer particles at island edges is not high enough to allow frequent upward jumps and to surmount the downward flux.
The result is consistent with a previous study of the CV model at small coverages and $R\gtrsim{10}^4$ \citep{carrascoPRE2023}.
As $R$ ($=\Rs$) increases, we observe the convergence to the ideal LBL case in which $\OP\to1$.

When $\DE$ increases ($\nu$ increases) and $R$ is kept constant, the diffusivity on the substrate is enhanced while the diffusivity on the film does not change.
A transition to ISL growth occurs in the two models, with $\OP$ showing very similar behavior.
The transition becomes sharper and the curves of the two models get closer in the limit $R\to\infty$.
In these conditions of slow deposition, the collective diffusion becomes increasingly relevant in comparison with the particle flux [see Eq. (\ref{defR})], so it more rapidly brings the system to equilibrium configurations.
These configurations are the same in the two models, but their paths to the equilibration are different, so for reaching similar values of $\OP$ it is actually necessary that the relaxation is fast.
These observations indicate the existence of a (quasi--)equilibrium wetting transition.
Such a view on equilibrium wetting is usually not taken in the literature, and it appears to be  cumbersome to study equilibrium wetting in that way, since already the results of Fig.~\ref{fig:comparison} indicate that the approach to the critical value $\Delta E = 0$ of the equilibrium wetting transition is approached rather slowly with $R$ increasing.

The $\OP(\DE)$ curves in Fig.~\ref{fig:comparison} have an approximately tanh--like shape that conveniently allows to determine an inflexion point $\DEm$ and a width $w$ by fitting them with
\be
 \OP (\DE) = c \left[ \tanh \left( \frac{\DE- \DEm}{w} \right) -1 \right] .
\label{OPfit}
\ee
The width $w$ increases as $R$ decreases or as $\Eb$ increases.
Since $\OP$ is slightly below 1 in the LBL regime ($\DE \to 0$), the fitting constant $c$ is slightly larger than $-1/2$.

In Ref.~\cite{empting2021}, the energetic equivalent of $\DEm$ was considered a transition point and termed ``dynamic wetting transition point''.
At this point and after the deposition of one ML, approximately $1/4$ of the substrate is covered with islands of height $2$ and another $1/2$ with islands of height $1$.
The finite width $w$ of the transition is a consequence of the growth process and is weakly affected by finite--size errors \cite{empting2021}.

For the analysis of the island formation in the subsequent sections, it is physically reasonable to identify the transition point as the one where $\OP$ has reached a value close to zero and almost exclusively two--layer islands are present.
This can be  conveniently realized by the condition that $\DE=\Eb-\Esub$ exceeds the critical value
\be
\DEc=\DEm+2w ,
\label{DEc}
\ee
which leads to $\OP (\DE) \approx 0.02$ (for $c=-1/2$).
Eq. (\ref{ECV}) then gives the minimum value of the ratio of diffusivities on the substrate and on the film for obtaining islands with two layers:
\be
\nuc=\exp{\left( \DEc\right)} .
\label{nuc}
\ee
As $R$ increases and the curves of the two models become closer, we obtain $w\to 0$ and  $\DEc\to 0$ ($\nuc\to1$) {(see Fig. \ref{fig:comparison})}, as one would expect from the equilibrium wetting transition \citep{venables,burkeJPCM2009,empting2021}.

In the following, we will take these two observations as a starting point and derive a criterion for the LBL--ISL transition which should work best when a rather sharp transition is seen in $\OP$ as a function of $\DE$ (e.g. Fig.~\ref{fig:comparison} for $R\geq{10}^6$).
However, for a given $R$ and larger $\Eb$, the transition becomes wider, and the mechanism of two--layer island formation is also influenced by second--layer nucleation from deposited particles.
This is discussed in Appendix \ref{app:character}.

\subsection{Phase diagram of the LBL--ISL transition}
\label{phasediagram}

\begin{figure}[!h]
  \includegraphics[width=0.48\textwidth]{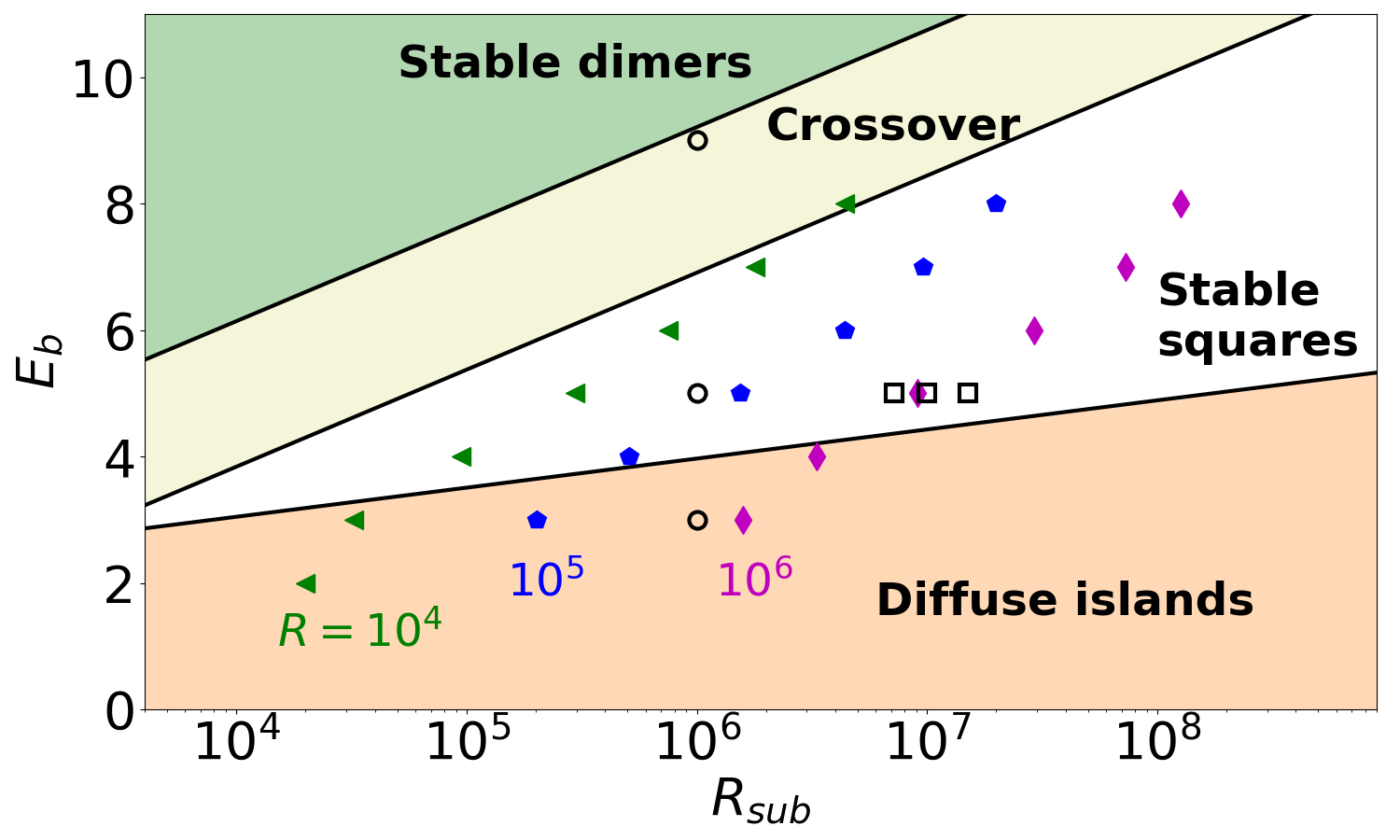}
  \caption{Transition points between ISL and LBL growth (shown as {filled} symbols) in the $R_\text{sub}$--$\Eb$ plane. For $\Eb$ fixed, ISL growth occurs to the right of the transition point.
  Thick lines denote the location of the scaling variables $Y_1=1$, $Y_1=10$, and $Y_2=0.3$, which roughly demarcate the {submonolayer} aggregation regimes (Sec. \ref{submonolayer}). Open circles {and squares indicate the growth parameters associated with the fluxes shown in Figs. \ref{fig:stepmoves} and \ref{fig:stepmoves1}, respectively}.
}
  \label{fig:Rsub_Eb}
\end{figure}

Using the order parameter data from simulations of the CV model, we have fit the $\OP (\DE)$ curves to Eq. (\ref{OPfit}) and determined the transition point $\DEc$ by Eq. (\ref{DEc}), from which $\nuc$ is obtained.
Figure \ref{fig:Rsub_Eb} shows the transition points in an $\Eb\times\Rs$ plane where the submonolayer growth regimes and their crossover regions are indicated.
Each colored full symbol is used to identify the data obtained with each diffusion--to--deposition ratio $R$ on the film.
For the physical interpretation of this diagram at constant bond energy $\Eb$, the variation of $R$ reflects a change in the base rate $k$ relative to the external flux and the variation of $\Rs$ reflects a change in the additional mobility of particles on the substrate [Eqs. (\ref{kCV}) and (\ref{defR})].
Most transition values are in region II of submonolayer growth and the lines of constant $R$ smoothly vary when crossing to region III. 

Transition values for constant $R$ and increasing $\Eb$ lie on tilted lines which reflect the increase in $\nuc$ for increasing $\Eb$.
In other words, increasing $\Eb$ with constant $R$ requires a larger enhancement of the substrate diffusivity (or larger energy difference $\DE=\Eb-\Esub$) for most islands to have two or more layers.
This suggests that there is no drastic change in the mechanisms that control this transition in the stable squares regime, where compact islands are formed. Although, there is a noticeable downward bending of the $R={10}^4$ line when it crosses between regimes III (diffuse islands) and II (stable squares), this downward bending is not evident for $R={10}^5$ and $R={10}^6$. The snapshots in Figs. \ref{fig:snap_sequence}(b) and \ref{fig:snap_sequence}(c) are obtained near the crossover between II and III.

In regime II, islands are compact and have flat borders, so that most aggregated particles have $2\text{--}4$ lateral neighbors \citep{oliveirareis2013}.
A particle that moves in the first layer and reaches a flat border has only one neighbor and may easily hop upwards, a process that is favored by high $\Rs$.
This particle and the particles directly deposited on the island top can execute random walks on a compact region and may nucleate a second layer island before hopping downwards.
In regime III,  particles with two lateral neighbors may also hop upwards; e.g kink particles at the irregular island borders. This increases the upward flux, leading to an even higher particle density on the second layer, thereby facilitating the nucleation of second-layer islands.

The formation of a second layer in the regime  I (stable dimers) is expected to be very difficult for two reasons: (i) particles in the first layer that reach the edge of an island irreversibly aggregate there (in the timescale of the layer formation); (ii) the islands in the first layer exhibit a DLA-like structure, so particles deposited directly onto the second layer often land on thin branches (thickness of approximately one site) from which they easily hop back to the first layer, where they irreversibly aggregate before nucleating a second layer island. Consequently, the transition line likely bends downward to avoid entering deeply into the stable dimer regime at much higher $\Rs$ than the ones we have simulated.
Thus, the second layer begins to be filled only for high coverages ($t$ close to $1$) in which the first layer islands have coalesced and the sites between their branches have been filled.

\subsection{Origin of particles in the second layer}
\label{subsec:origin}

In order to understand the source of the second layer particles, here we study the particle flux between different layers as the LBL-ISL transition is crossed.

\begin{figure}[!h]
  \includegraphics[width=0.45\textwidth]{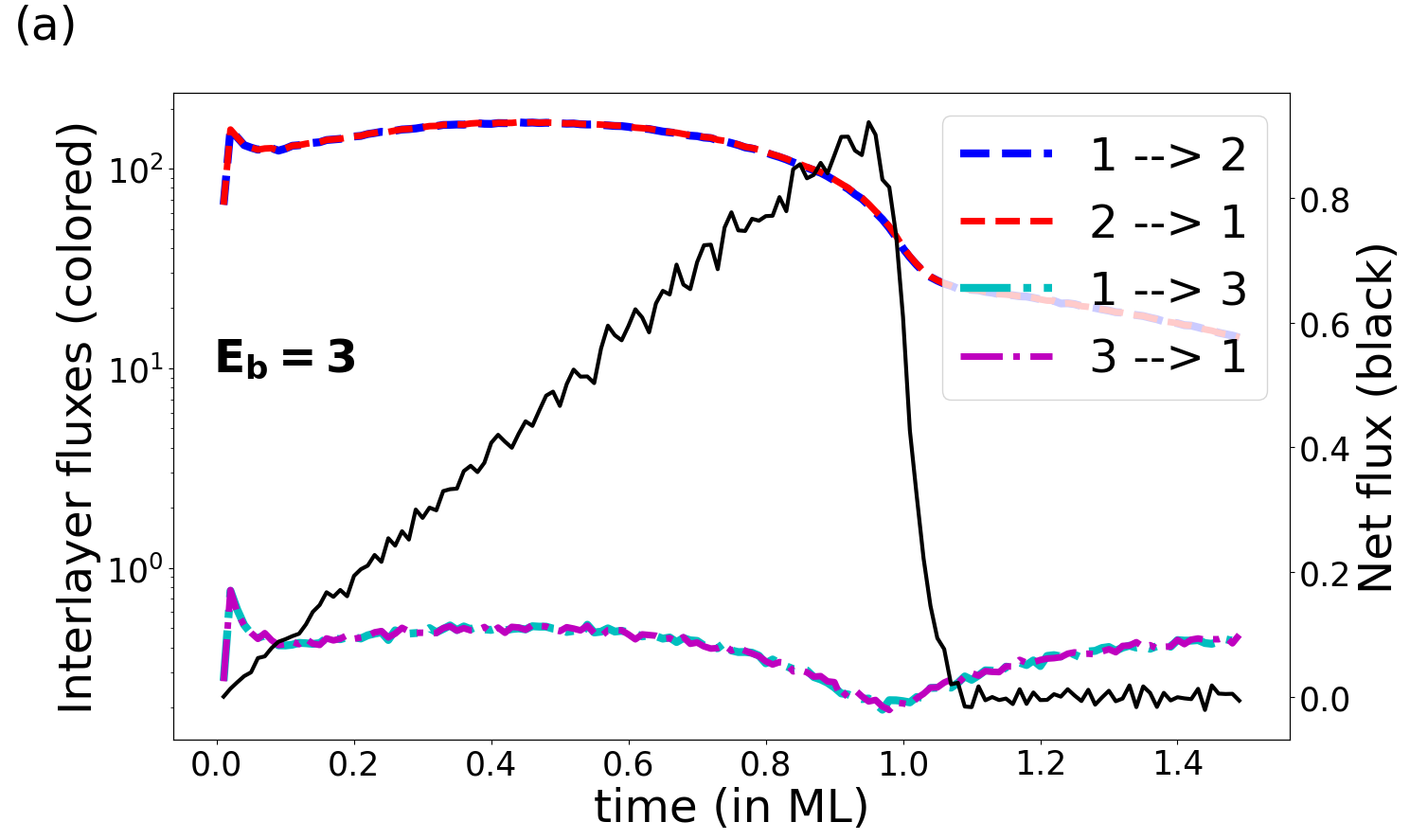}\\
  \includegraphics[width=0.45\textwidth]{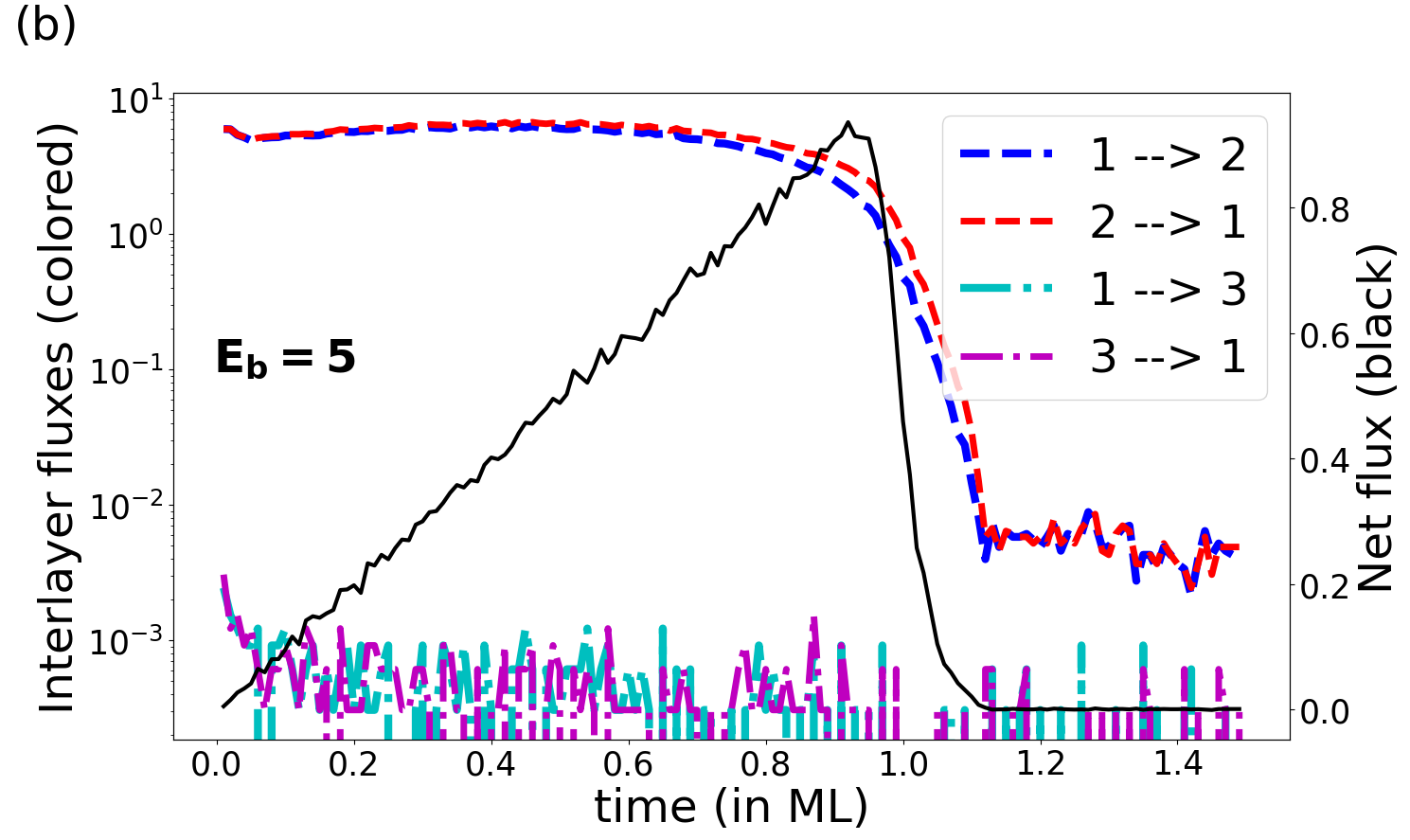}\\
  \includegraphics[width=0.45\textwidth]{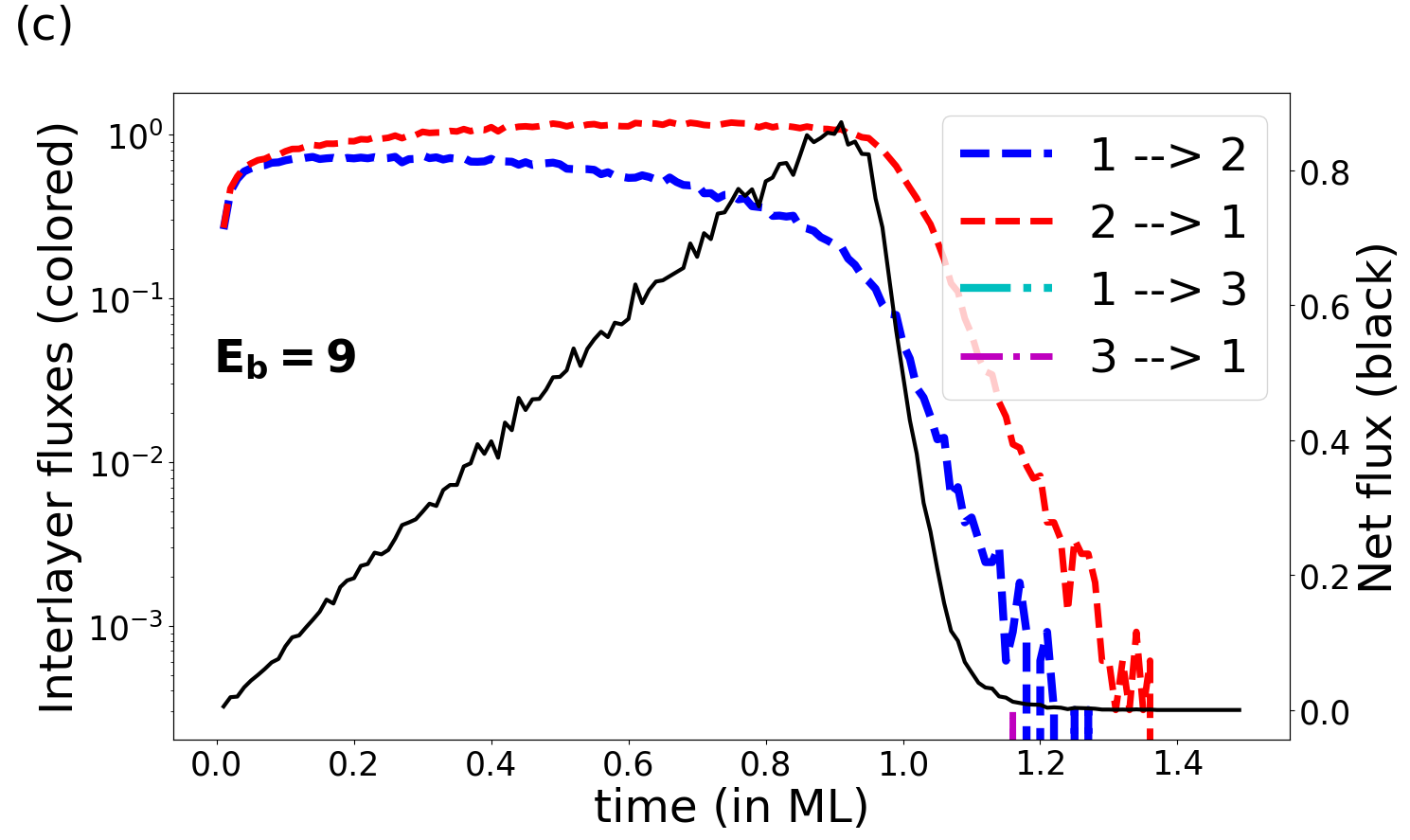}
  \caption{Layer exchanging fluxes in the CV model for $\Rs=R=10^6$ ($\nu=1$) and three different values of $\Eb$ (LBL conditions):
  (a) $\Eb=3$, diffuse island regime;
  (b) $\Eb=5$, stable square island regime;
  (c) $\Eb=9$, close to the stable dimer regime.
  Colored lines with a logarithmic flux axis to the left show the flux due to individual step moves (red: upward hops, blue: downward hops).
  Black lines with a linear flux axis to the right show the net flux down to the first layer.     }
  \label{fig:stepmoves}
\end{figure}

Beginning with the homoepitaxial growth case ($\nu=1$), Figs.~\ref{fig:stepmoves}(a)-(c) show the fluxes up to the deposition of $1.5$ layers for $\Rs=R={10}^6$ and three values of $\Eb$.
These model parameters are indicated in the phase diagram of Fig.~\ref{fig:Rsub_Eb} (open circles), showing that the submonolayer regimes from diffuse islands (III) via stable square islands (II) to stable dimers (I) are covered.
The net flux $\Phi_{\text{net},1}$ [Eq. (\ref{Phinet})] is close to the prediction of Eq. (\ref{LBLflux}) in all cases, which shows that the growth is close to the ideal LBL.

For $\Eb=3$, Fig.~\ref{fig:stepmoves}(a) shows that the fluxes $\Phi_{2\to 1}$ and $\Phi_{1\to 2}$ are the dominant ones; the fluxes due to direct hops $1 \leftrightarrow 3$ and $1 \leftrightarrow 4$ (not shown) are comparably negligible.
Moreover, $\Phi_{2\to 1}$ and $\Phi_{1\to 2}$ are larger than the net flux to the substrate by factors $\gtrsim{10}^2$ even after the deposition of $1$~ML.
Fig. \ref{fig:comparison} shows that these conditions are close to the sharp LBL-ISL transition that occurs for a small value $\nu<2$ ($\DE<0.5$).
For $\Eb=5$ [Fig.~\ref{fig:stepmoves}(b)], the individual fluxes of $1 \leftrightarrow 2$ hops are still larger than the net flux to the substrate by a factor of order $10$; they are also much larger than the fluxes to and from layer 3.

The high values of $\Phi_{2\to 1}$ and $\Phi_{1\to 2}$ compared to $\Phi_{\text{net},1}$ suggest that the LBL growth evolves in quasi-equilibrium conditions, i.e. with almost balanced particle exchange between the layers $1$ and $2$.
This is consistent with the similar curves for two relaxation models in Fig. \ref{fig:comparison} ($\Eb=3$).

Figure \ref{fig:stepmoves}(c) shows a large decrease in the fluxes $1 \leftrightarrow 2$ when $\Eb$ is increased to $9$.
The fluxes $\Phi_{2\to 1}$ and $\Phi_{1\to 2}$ are significantly different and they are on the same order of magnitude of the net flux $\Phi_{\text{net},1}$.
However, the condition $\nu=1$ is far from the transition point for such a large value of $\Eb$; Fig. \ref{fig:Rsub_Eb} shows that the LBL-ISL transition occurs for $\nu>{10}^2$.

\begin{figure}[!h]
	\includegraphics[width=0.45\textwidth]{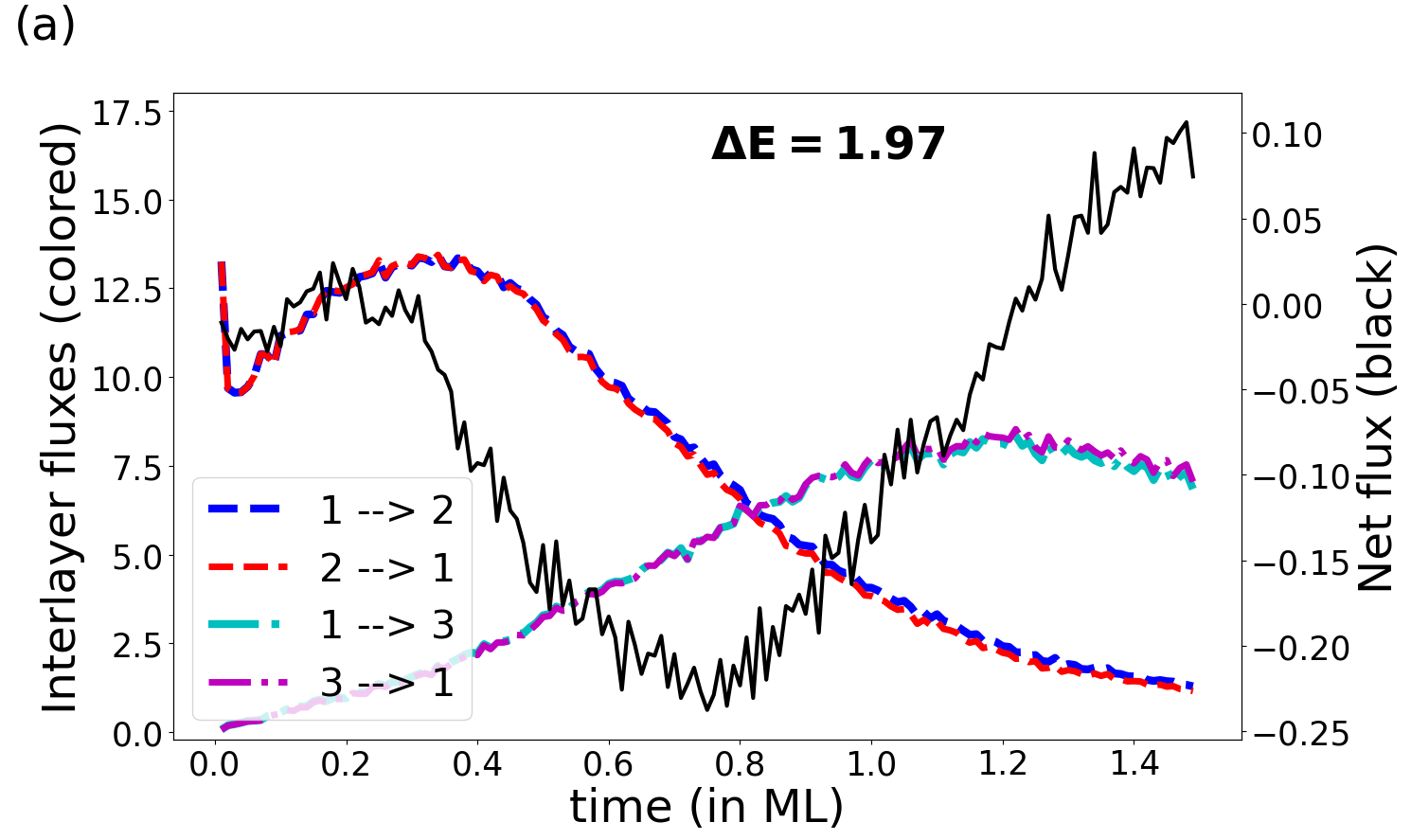}\\
	\includegraphics[width=0.45\textwidth]{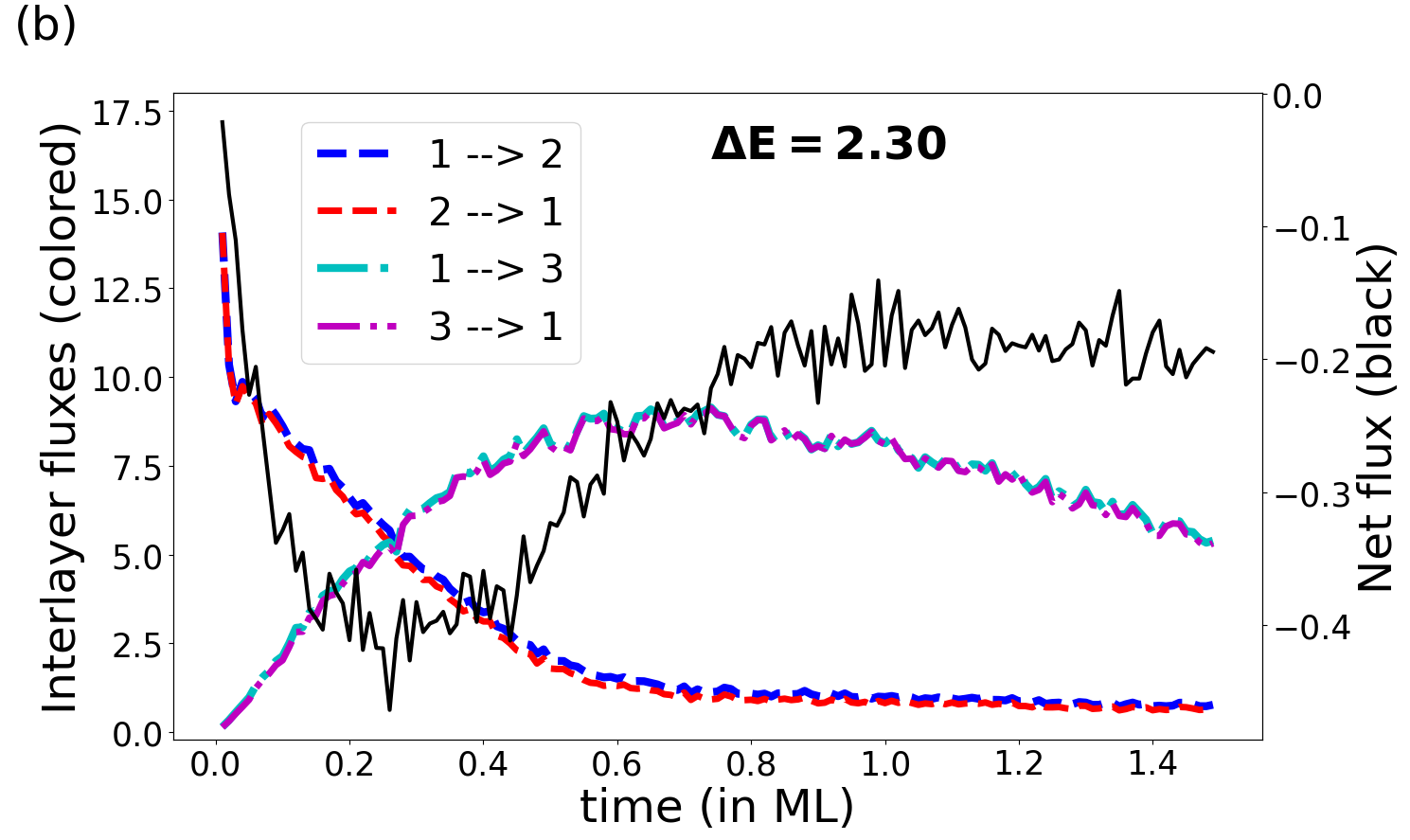}\\
	\includegraphics[width=0.45\textwidth]{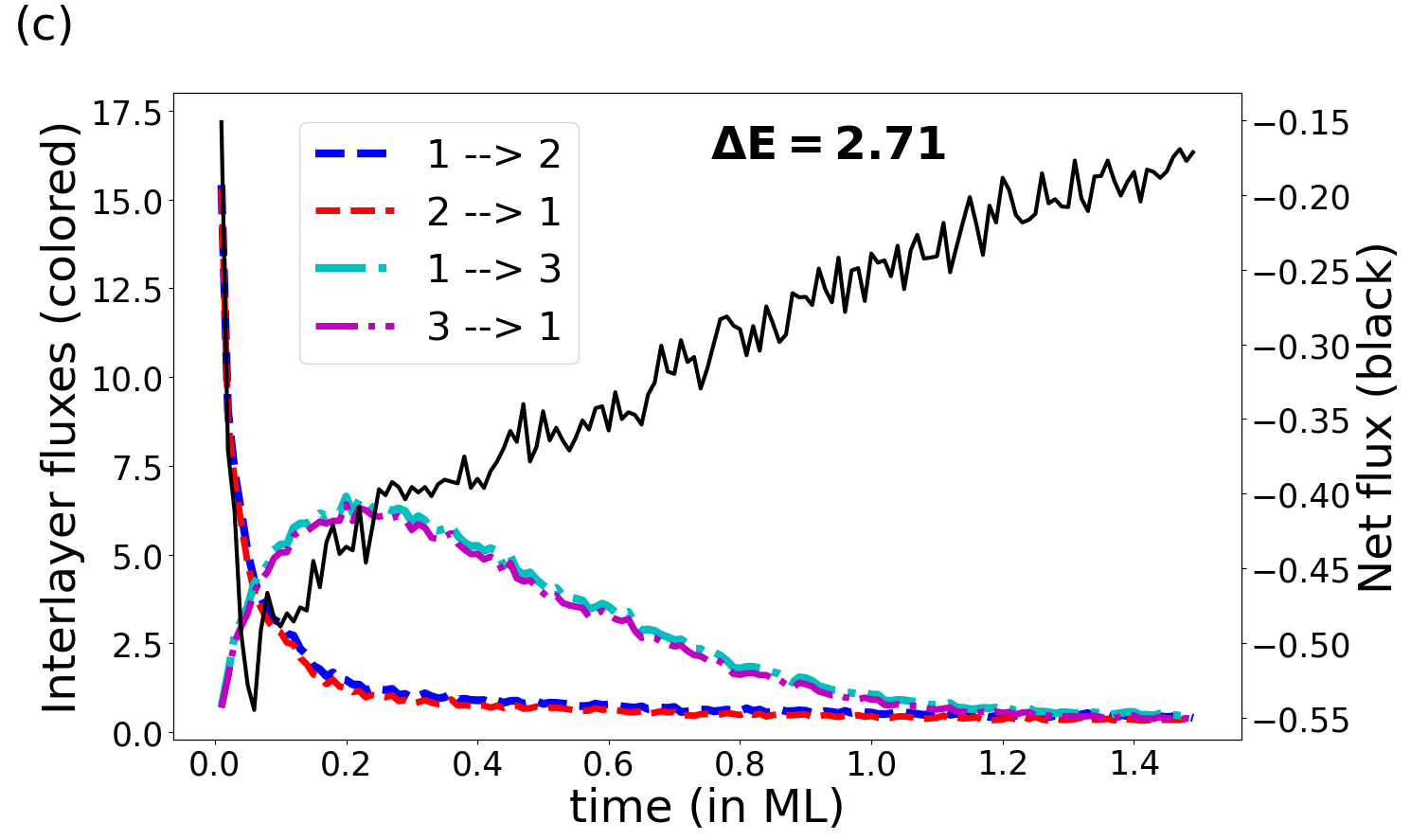}
	\caption{Layer exchanging fluxes in the CV model of heteroepitaxy for $\Eb=5$, $R=10^6$, and:
  (a) $\nu=7.196$ $(\Delta E=1.97)$;
  (b) $\nu=10$ $(\Delta E=2.30)$;
  (c) $\nu=15$ $(\Delta E=2.71)$.
  Colored lines with a logarithmic flux axis to the left show the flux due to individual step moves (red: upward hops, blue: downward hops).
  Black lines with a linear flux axis to the right show the net flux down to the first layer.
}
  \label{fig:stepmoves1}
\end{figure}

Now we analyze the fluxes near the transition or in the ISL phase for $\Eb=5$ and $R={10}^6$.
Figures \ref{fig:stepmoves1}(a)-(c) show the fluxes for three values of $\nu>1$ ($\DE>0$), which may be compared with Fig. \ref{fig:stepmoves}(b) for $\nu=1$ ($\DE=0$).
The net flux evolution is very different from the prediction of ideal LBL growth [Eq. (\ref{LBLflux})];
the negative values up to $t=1$ are consistent with dominant particle flux to the layers $z\geq2$.
However, the fluxes between $z=1$ and another layer $i$, $\Phi_{i\to 1}$ and $\Phi_{1\to i}$, have small differences for all $i\geq2$.

When $\nu=7.196$ $(\Delta E=1.97)$, an upward flux larger than that of $\nu=1$ is expected, which is consistent with the change of $\Phi_{1\to 2}$ by a factor $\approx2$ from Fig. \ref{fig:stepmoves}(b) to Fig. \ref{fig:stepmoves1}(a).
The parameters that control the diffusion at $z=2$ ($R$ and $\Eb$) were kept constant, so they do not contribute directly to a change in $\Phi_{2\to 1}$.
However, $\Phi_{2\to 1}$ is still very close to $\Phi_{1\to 2}$, which means that the increase of $\Phi_{2\to 1}$ was caused by the increase of the particle density in the second layer as a response to the increase in the upward flux.
At $t\gtrsim0.8$, Fig. \ref{fig:stepmoves1}(a) shows that the fluxes between the layers $1$ and $3$ become dominant, but they are also much larger than their difference.

When $\nu=10$ $(\Delta E=2.30)$, which is a value close to the transition point ($\nu=9.68$), Fig. \ref{fig:stepmoves1}(b) shows that $\Phi_{1\to 2}\approx\Phi_{2\to 1}\gg\Phi_{\text{net},1}$ until $t\approx0.3$, in which the exchange between the layers $1$ and $2$ is dominant.
At longer times, the exchange between layers $1$ and $3$ dominates, but $\Phi_{1\to 3}$ and $\Phi_{3\to 1}$ are also much larger than the difference between them.
Fig. \ref{fig:stepmoves1}(c) shows similar results for $\nu=15$ $(\Delta E=2.71)$, which is well above the transition point.
For $t\gtrsim0.5$, the fluxes between layers $1$ and $4$ become dominant but are also close to each other, which indicates that islands with several layers are formed.

The above results imply that quasi-equilibrium conditions remain near and above the LBL-ISL transition.
Moreover, in all cases the dominant fluxes involving the layer $1$ ($\Phi_{1\to i}$ and $\Phi_{i\to 1}$, for $i\geq 2$) are larger than the deposition flux (set to $1$ in the model).
This means that the particles forming islands on the second layer mostly originate from the high flux from the island borders, not from the deposition flux.
Consistently, Figs.~\ref{fig:snap_sequence}(a)-(c) show that most second layer islands nucleate near the borders of first layer islands.

\section{Thermodynamic approach to the LBL--ISL transition}
\label{thermodynamic}

\subsection{Dynamic wetting transition}
\label{sec:dynwet}

In the initial stage of submonolayer growth, an initial island density $\Nisl$ is quickly established which then slowly changes. Hops at the edge of these islands populate the second layer on top of the island. We assume that these hops lead to an equilibrated island within a short time, and we describe this with a simple thermodynamic model.
These assumptions are reasonable given the quasi-equilibrium conditions near the LBL-ISL transition shown in the above simulations.

Consider an island with a square shape and with the same bond energies of the CV and DM models.
Suppose this island has $N=N_1+N_2$ particles, with $N_1$ in the first layer and $N_2$ in the second layer.
The total binding energy of the particles in the first layer is
\be
 E_1 \approx -N_1 (2 \Eb + \Esub) + 2 \sqrt{N_1} \Eb
\ee
Here, $-N_1\,2\Eb$ is the (bulk) binding energy due to in--plane bonds (each particle has 4 bonds but shares each bond with another particle), $-N_1 \Esub$ is the binding energy with the substrate, and $2 \sqrt{N_1} \Eb$ is the line energy contribution of the missing bonds of the edge particles ($4\sqrt{N_1}$ is the number of edge particles for a square island and each missing bond counts $1/2$). Analogously the total binding energy of the particles in the second layer is
\be
 E_2 \approx -N_2\, 3 \Eb  + 2 \sqrt{N_2} \Eb
\ee
where $\Esub$ has been replaced by $\Eb$ since the particles are bonded to the film particles below.

We consider the normalized energy sum $\varepsilon(N_1)=E/\Eb$ as a function of $N_1$ under the constraint that $N$ is fixed:
\be
\begin{split}
  \varepsilon(N_1) = & -3 N + N_1 \left(1 - \frac{\Esub}{\Eb} \right) \\
  & + 2\left( \sqrt{N_1}+ \sqrt{N-N_1}\right) \;.
\end{split}
\ee
$\varepsilon(N_1)$ is defined for $N_1$ between $N/2$ (island with height 2 with equal number of particles in both layers) and $N$ (island of height 1).
This function has a single maximum between $N/2$ and $N$, and thus the minimal energy configuration is either $N_1=N/2$ or $N_1=N$.
This explains the sharp transition between almost ideal LBL ($\OP=1$) and nearly complete filling of the second layer ($\OP=0$) observed in our simulations with large $R$ (Fig. \ref{fig:comparison}).
There is a critical point $N=\Nc$ in which the energies of those configurations are equal:
\be 
 \Nc =\alpha {\left(\frac{\Eb}{\ln{\nu}}\right)}^2 .
\label{defNc}
\ee 
with $\alpha = 16 (\sqrt{2}-1)^2 \approx 2.75$.
For $N<\Nc$ the monolayer island ($N_1=N$) is stable, while for $N>\Nc$ the island of height $2$ is stable ($N_1=N/2$).

In homoepitaxial growth ($\nu=1$), the critical monolayer island size $\Nc$ diverges, so all finite first islands are energetically stable against being covered by a second layer of particles.
This is a condition of equilibrium wetting, in which the particle-substrate and particle-film interactions are equal.
The result parallels our simulations with $\nu=1$ that always show nearly LBL growth.

Now consider the heteroepitaxial case of $\nu>1$ ($\Delta E > 0$).
Small islands with sizes $N<\Nc$ will occupy only the first layer.
However, islands with two layers may be formed when the growing submonolayer islands reach the value $\Nc$.
Conditions for this  can be formulated using submonolayer growth theories and allow us to predict whether the transition can occur or not, as anticipated by the numerical work.
The connections with those theories are presented in the following subsections.

\subsection{Connection of dynamic transition and submonolayer growth}
\label{sec:quantitative}

As a first approximation, LBL growth is expected to persist if the submonolayer islands merge before they reach the critical size $\Nc$ for second layer formation.
The maximum size of the island before merging can be estimated as the reciprocal of the island density, $1/\Nisl$, which may be obtained theoretically or in simulations.
Thus, if $\Nc<1/\Nisl$, the second layer is formed, but this is not possible if $\Nc>1/\Nisl$, so
the condition for the LBL-ISL transition is
\be
\Nc\left(\nuc\right) \approx 1/\Nisl .
\label{scalingNc}
\ee
This sets the minimal value of $\nu$ for second layer formation.

The scalings of the island density $\Nisl$ in the submonolayer regimes I and II of the CV model fit the results of rate equation theories with maximal unstable island sizes $i^*=1$ and $i^*=3$, respectively:
\be 
  \Nisl \sim t^{\frac{1}{i^*+2}}\; \exp\left(-\frac{\Estar}{i^*+2}\right) \; \Rs^{-\frac{i^*}{i^*+2}}\;,
  \label{eq:nisl}
\ee 
where $\Estar<0$ is the binding energy associated with that maximal unstable size: $\Estar=0$ for $i^*=1$ (regime I of stable dimers) and $\Estar=-2\Eb$ for $i^*=3$ (regime II of stable $2\times2$ squares).
The time dependence in Eq. (\ref{eq:nisl}) is weak, so we consider a quasi-stationary island density and only consider the dependence on $\Rs$ and $\Estar$ further.

Fig. \ref{fig:Rsub_Eb} shows that most of the transitions analyzed here are in regime II.
Moreover, the previous thermodynamic approach assumed square island shapes that are characteristic of regime II.
Thus our comparisons between theoretical and numerical results are performed in this regime. 
Substituting the critical size of the thermodynamic approach [Eq. (\ref{defNc})] and the relation of rate equation theory for $i^*=3$ [Eq. (\ref{eq:nisl})] in Eq. (\ref{scalingNc}), we obtain
\bea 
  \frac{\ln{\nuc}}{\Eb} \sim \exp( \Eb/5 )\,  {\left(\nuc R\right)}^{-3/10} . 
    \label{eq:esubc1}
\eea
Rearranging gives
\be
  \underbrace{\frac{10}{3} \ln \ln \nuc + \ln \nuc}_{y_\nu} = \underbrace{\frac{10}{3} \ln \Eb + \frac{2}{3} \Eb}_{y_E} - \ln R \;,
  \label{eq:transition_y}
\ee
which defines the scaling variables $y_\nu$ related to the increased diffusivity on the substrate and $y_E$ related to the particle-particle interaction energy.
The double logarithmic terms in Eq.~(\ref{eq:transition_y}) could be neglected for $\ln \nuc \gtrsim 8$ and $\Eb \gtrsim 15$, but these ranges are outside the range of our numerical data, which justifies the inclusion of those terms in the definition of the scaling variables.

\begin{figure}[!h]
  \includegraphics[width=0.48\textwidth]{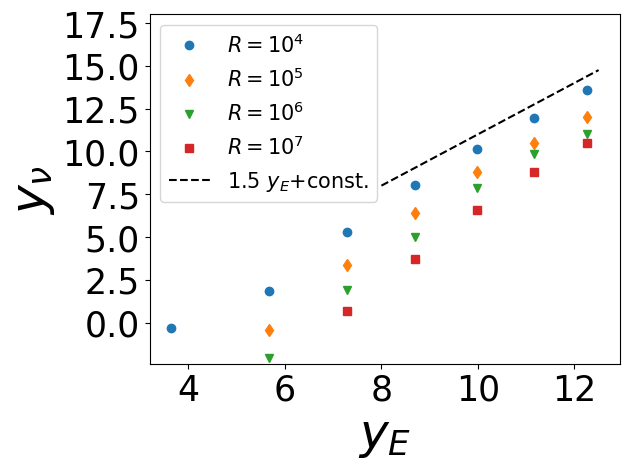}\\
  \includegraphics[width=0.48\textwidth]{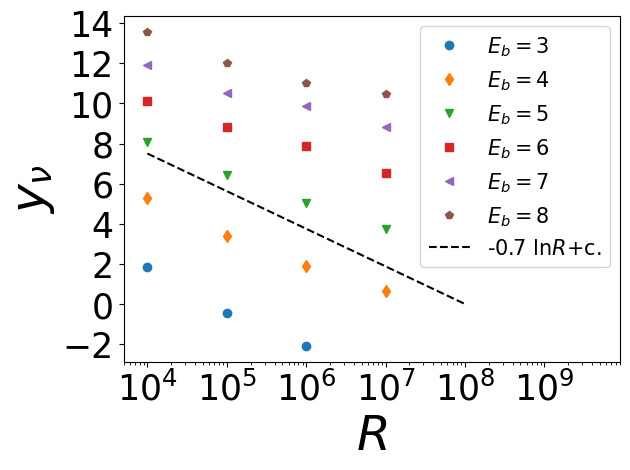}
	\caption{Scaling variable $y_\nu$ as a function of (a) the scaling variable $y_E$ and (b) $R$ from CV model simulations. The scaling variables are defined in Eq.~(\ref{eq:transition_y}).
	}
	\label{fig:ys_R}
\end{figure}

Fig.~\ref{fig:ys_R}(a) shows $y_\nu$ versus $y_E$ for several values of $R$.
It qualitatively confirms the simple linear dependency predicted by Eq.~(\ref{eq:transition_y}) for constant $R$, but with slopes $\sim1.5$ that differ from the predicted value $1$.
Fig.~\ref{fig:ys_R}(b) shows $y_\nu$ versus $R$ for several values of $\Eb$, which also confirms a linear dependence on $\ln R$ (for constant $\Eb$) but with a slope $\approx -0.7$ differing from the predicted value $-1$.

Despite some quantitative discrepancies, the thermodynamic approach captures the main controls of the transition observed in the simulations: the minimal ratio of substrate and film diffusion coefficients for second layer nucleation, $\nuc$, increases as the bond energy $\Eb$ increases and decreases as the diffusion-to-deposition ratio $R$ increases.
The phase diagram in Fig. \ref{fig:Rsub_Eb} shows that these trends extend to regime III of islands with diffuse borders and to the crossover with regime I in which the islands become branched.
These physical interpretations may guide the choice of substrate and operation temperature for the deposition of a given material with the aim of obtaining LBL or ISL growth.

\subsection{Improved connection of dynamic transition and submonolayer growth}
\label{improved}

The movies ESM1 and ESM2 in the Electronic Supplement show the evolution of the deposits grown near the transition points with $R={10}^5$ and $R={10}^6$, respectively, for $\Eb=5$.
These movies and the snapshots in Figs.~\ref{fig:snap_sequence}(a)-(c) reveal that two--layer island formation does not occur at the point where first--layer islands merge, i.e. the condition $N_c \approx \Nisl^{-1}$ is not met at the transition.
Rather, the formation of two--layer islands occurs when the first layer islands are still well separated, and for increasing $\Eb$ the maximum size of islands covered with a second layer becomes smaller.

The islands mainly form by free monomers on the substrate which jump to the second layer, so enough monomers (free particles) must be present around the first layer islands.
For this reason, we must consider explicitly the time--dependence of the monomer and the island density.
Each growing first layer island 
has a capture zone with area $\Nisl(t)^{-1}$ and, when this island is still small, it is surrounded by $\Nm(t)/\Nisl(t)$ monomers, where $\Nm(t)$ is the free particle density.
Initially, this number is much larger than the number of particles in the first layer island, but at some critical time (or coverage) $t_c$ these numbers become equal.
We regard this as an improved estimate of the time frame within which two--layer island formation becomes possible.

The average number of particles in a first layer island in these conditions will be $a t_c/\Nisl(t)$, where $a$ is a factor of $O(1)$ ($a=1/2$ in the stable dimer regime, otherwise somewhat smaller).
The condition for second layer formation then becomes
\be 
\begin{split}
  & \Nm(t_c) \; \Nisl^{-1}(t_c)  \sim  t_c\; \Nisl^{-1}(t_c) \\
  & \Rightarrow \Nm(t_c) \sim t_c\;.
\end{split}
\ee 
Using the scaling expression for the monomer density, $\Nm \sim t^{-1/(i^*+2)}\, \exp(E^*/(i^*+2))\, \Rs^{-2/(i^*+2)}$ \citep{venables1987,barteltSSL1995,oliveirareis2013}, and solving for $t_c$ in regime II (stable 2x2 squares, $i^*=3$), we obtain
\be
 t_c \sim  \exp( -\Eb/3 )\,  \Rs^{-1/3} .
   \label{eq:thetac}
\ee

For the LBL--ISL transition point, we retain the condition that the number of particles in the first layer island equals the critical number $N_c$ (Eq.~(\ref{defNc})).
However, the condition from Eq.~(\ref{scalingNc}) is now slightly modified to
\be
  N_c \approx  t_c \; \Nisl^{-1}(t_c) ,
\ee
which accounts for the (possibly small) fraction of the area of the capture zone occupied by the island.
Using Eq.~(\ref{eq:thetac}) for $t_c$ and Eq.~(\ref{eq:nisl}) for $\Nisl$ with $i^*=3$, we finally obtain
\be
  \frac{\ln\nuc}{\Eb} \left( = \frac{\Delta E_c}{\Eb} \right) \sim \exp( \Eb/3 )\,  \Rs^{-1/6}  \;.
\label{improvedscaling}
\ee

\begin{figure}[!h]
	\centerline{\includegraphics[width=0.48\textwidth]{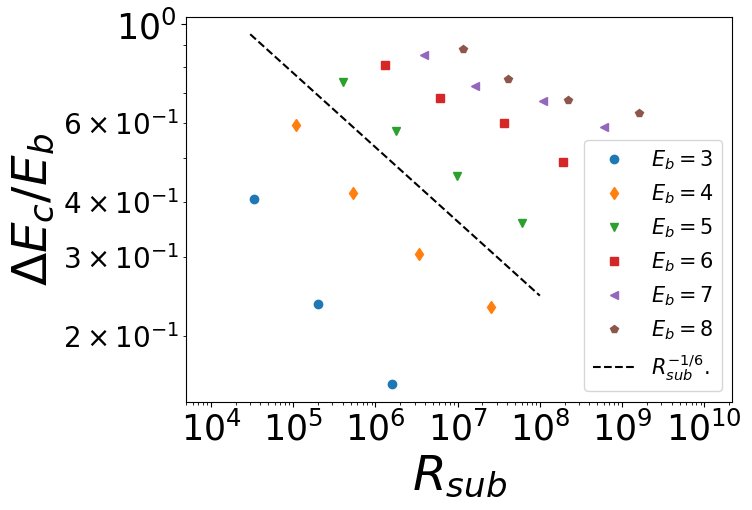}}
	\caption{$\Delta E_c/\Eb$ as a function of $\Rs$ for various $\Eb$ values from the CV model simulations.
	The dashed line with slope $-1/6$ is the prediction from scaling theory.}
  \label{fig:esubc}
\end{figure}

Figure \ref{fig:esubc} shows $\ln \nuc/\Eb=\Delta E_c/\Eb$ as a function of $\Rs$ from CV model simulations for various $\Eb$.
The $R^{-1/6}$ behavior from the refined version of scaling theory [Eq. (\ref{improvedscaling})] is consistent with the data for $\Eb=3$, $4$, and $5$.
For higher $\Eb=5$, the decay of $\ln \nuc/\Eb$ with $\Rs$ becomes weaker.


Overall, this exemplifies the extremely slow approach to the equilibrium wetting transition value
$\nuc=1$ ($\Delta E_c=0$) which practically cannot be reached in an experimental setup (where typically $R \sim 10^{8} \dots 10^{10}$ and $\Eb \sim 5 \dots 20$).
Thus, according to this analysis, the dynamic wetting transition from ISL to LBL is always occurring in the equilibrium regime of partial wetting, which entails the possibility of strong, dewetting post--growth effects.

\section{Summary and conclusions}
\label{conclusion}

We have analyzed the mechanism of island formation and the transition between island and layer--by--layer growth for heteroepitaxial growth on 
substrates with a binding energy $\Esub$ to film molecules different from the binding energy $\Eb$ between the monomers themselves. 
The analysis has been carried out using a minimal lattice  model which features an enhanced diffusion constant for particles on the substrate compared
to particles on the film. The ratio of these diffusion constants $\nu$ can be connected to the binding energy difference $\Delta E = \Eb-\Esub$ 
through a detailed balance argument.

Nucleation of islands with at least two layers of particles is dominated by an enhanced hopping rate between layers 1 and 2 for {$\nu\geq\nuc>1$} ($\Delta E\geq\Delta E_c>0$), signifying a weak {substrate-particle interaction}.
{Comparably, in equilibrium conditions, two--layer islands appear for $\nu$ only slightly above $1$.
Near the ISL--LBL transition of the heteroepitaxial deposition models, the particle fluxes between two different layers greatly exceed the net flux between them, indicating that the transition occurs in quasi-equilibrium conditions.}
The stability of the islands was {then} discussed with an equilibrium argument, comparing the energy of 1--layer islands and 
2--layer islands with the same number of particles, leading to a minimum 2--layer island size which is stable. This condition has been linked
to well--known scaling expressions for the submonolayer island density and gives a scaling expression for the ISL--LBL transition point $\Delta E_c$.

Comparison with KMC simulation results shows that the new scaling theory explains the trends in the  dependence of $\Delta E_c$ on $\Eb$ and the
ratio $R$ of diffusion to flux very well, with some quantitative deviation in numerical coefficients. An improved version of the
scaling theory incorporates the condition that 2--layer islands can only be formed by attachment to and jumps of free monomers on top of an island 
in the capture zone of this island and predicts a very slow approach to the equilibrium wetting transition $\Delta E_c \propto R^{-1/6}$.
Therefore, within this model, the ISL--LBL transition in film growth is very different from the equilibrium wetting transition and always occurs in the partial wetting regime.

Although the model is minimal and has been exemplified with simulations on a simple cubic lattice, the assumptions are quite generic and
we believe that our results can be generalized to other models of thin--film growth. The nucleation mechanism of 2--layer islands via
hopping from the substrate has not, to the best of our knowledge, been systematically analyzed before, although the influence of weaker substrate binding on hopping rates has been effectively used previously in the simulation description of island formation, as e.g. in Refs.~\cite{luPRMat2018,gervilla2020} for the case of silver on a weak substrate or in Refs.~\cite{koerner2011,janke2021} for the case of C60 growth on CaF$_2$.


\begin{acknowledgments}

The authors are grateful for the support of mutual research visits by the German Academic Exchange Service (DAAD) with funds from the German Federal Ministry of Education and Research (BMBF) and by the Brazilian agency CAPES (project 88881.700849/2022-01), all within the joint DAAD-CAPES project "MorphOrganic".   

FDAAR acknowledges support from the Brazilian agencies CNPq (305570/2022-6) and FAPERJ (E-26/210.040/2020, E-26/201.050/2022, E-26/210.629/2023).

ISSC acknowledges support from FAPDF (00193-00001817/2023-43).

MO thanks Frank Schreiber for stimulating discussions.

\end{acknowledgments}

\section*{Author contributions}
\label{contributions}


Conceptualization: FDAAR, MO;
methodology: FDAAR, MO, ISSC;
software: FM, ISSC;
validation: all authors;
formal analysis: MO, FDAAR;
investigation: all authors;
resources: MO, FDAAR;
data curation: FM, CCL, MO;
writing---original draft preparation: MO;
writing---review and editing: all authors;
visualization: FM, CCL, ISSC;
supervision: MO;
project administration: MO, FDAAR;
funding acquisition: MO, FDAAR.

\begin{appendix}

\section{Character of the LBL-ISL transition for larger $\Eb$}
\label{app:character}

\begin{figure}[!h]
 \includegraphics[width=0.48\textwidth]{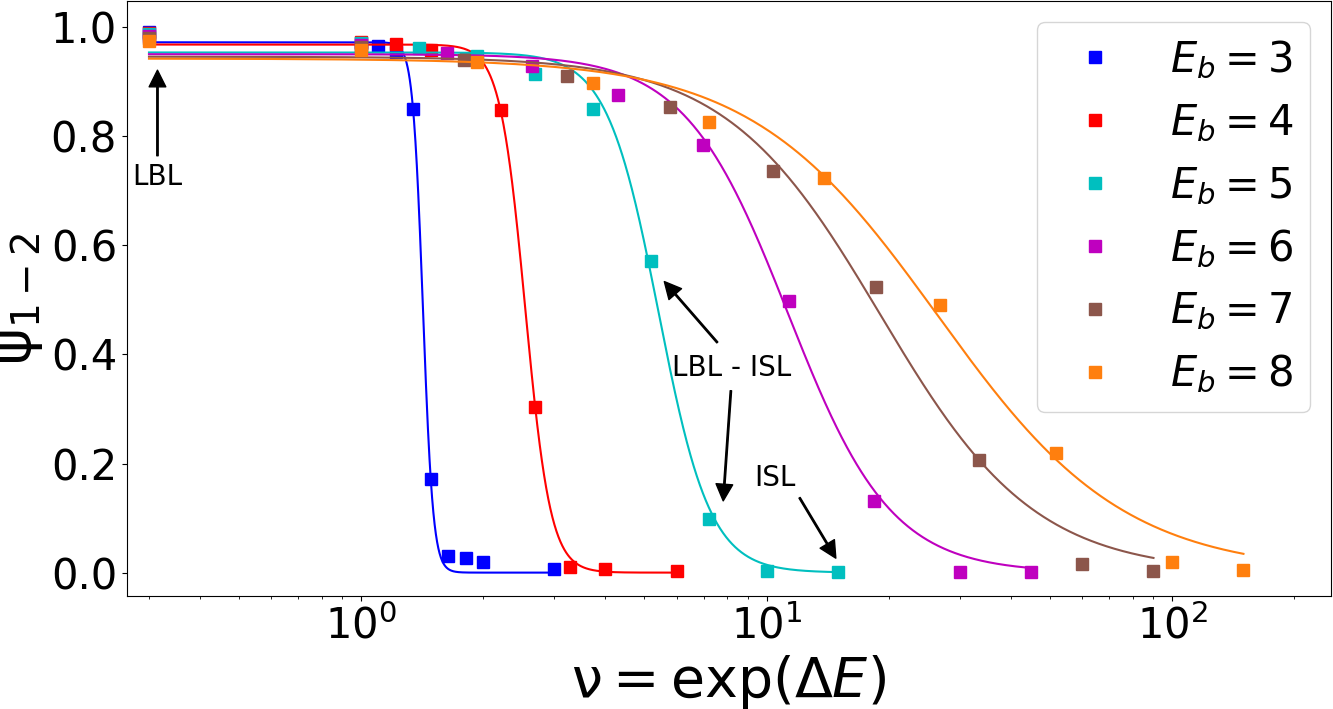}
  \caption{ Order parameter $\OP$ as function of $\nu=\exp(\Delta E)$ for $R=10^6$ in the CV model. Data points and fits (full lines) as shown here correspond to the ISL-LBL transition values $\Rs=R\nuc$ shown in Fig.~\ref{fig:Rsub_Eb} (for $R=10^6$). The arrows indicate the values of $\nu$ corresponding to the snapshots shown in Fig.\ref{fig:eb5_snaps1} and Fig.\ref{fig:eb5_snaps2}. The sequence begins with $\nus = 1$ ($\ln \nus = 0$), where LBL is clearly observed. It goes on through the transition regime at $\nus = 5.18$ and $\nus = 7.20$, and concludes in ISL at $\nus = 15$.}
  \label{fig:op_R6}
\end{figure}


The scaling theory of Sec.~\ref{sec:dynwet} for the ISL--LBL transition assumes a quick formation of equilibrium two--layer islands in the early stages of submonolayer growth.
In the theory, the transition is sharp, so it should work best for low $\Eb$ and large $R$ where the width of the transition is small, see Fig.~\ref{fig:op_R6} for the order parameter behavior for fixed
$R=10^6$ and various $\Eb$; for $\Eb=3$ the transition is almost sharp.
For larger $\Eb$, the width of the transition quickly becomes large, but the scaling theory does not give an explanation for this increased width.
The inspection of snapshots for such conditions suggests that the formation of equilibrated islands competes with another mechanism in the transition region.

\begin{figure*}[t]
  \includegraphics[width=\textwidth]{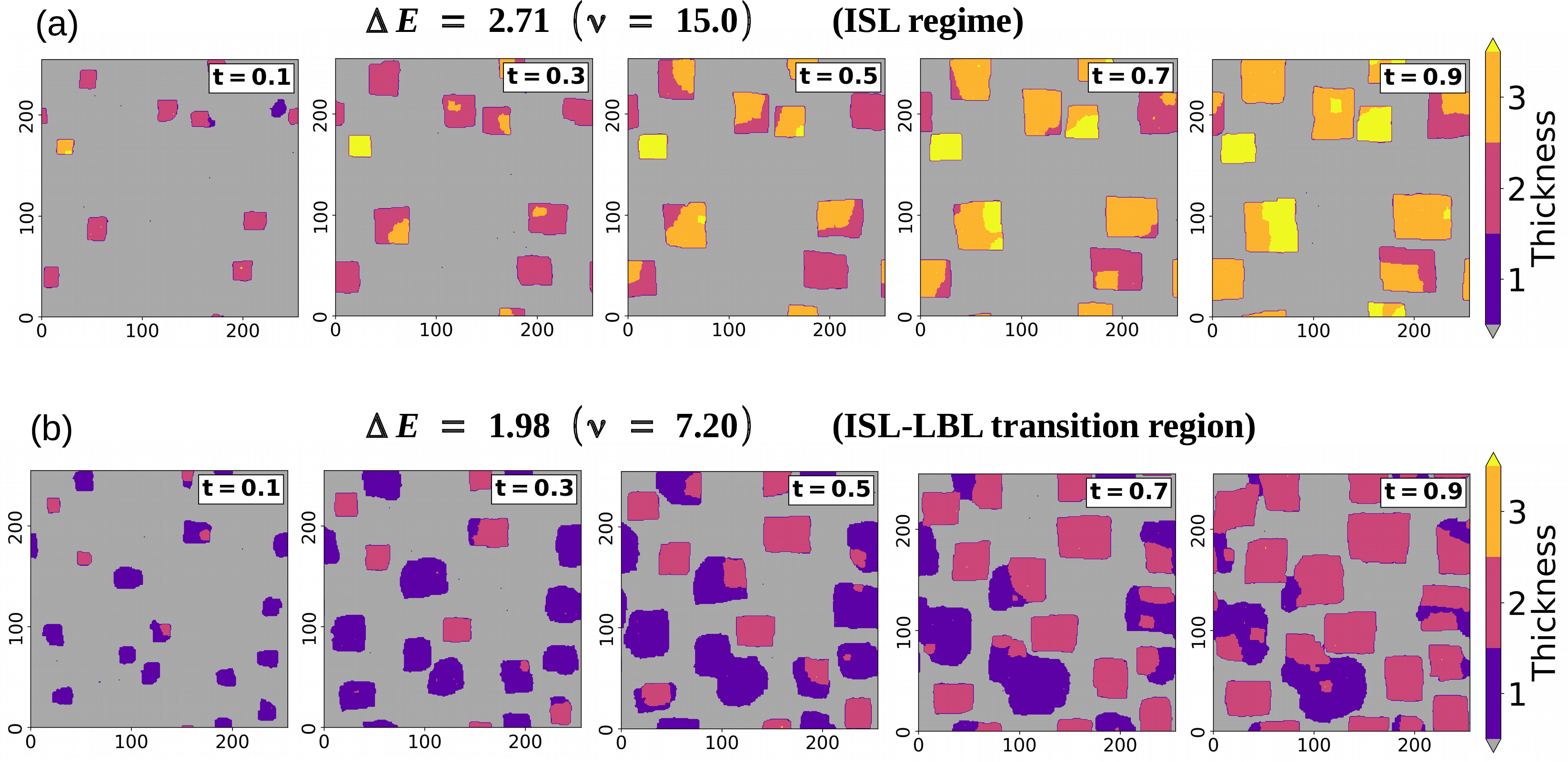}
   \caption{ Sequence of snapshots for $R=10^6$, $\Eb=5$. Upper panel:  $\Delta E=2.71$ ($\nus=15$, ISL regime). Lower panel  $\Delta E=1.98$ ($\nus=7.20$, beginning of ISL--LBL transition region).  }
   \label{fig:eb5_snaps1}
 \end{figure*}

\begin{figure*}
 \includegraphics[width=\textwidth]{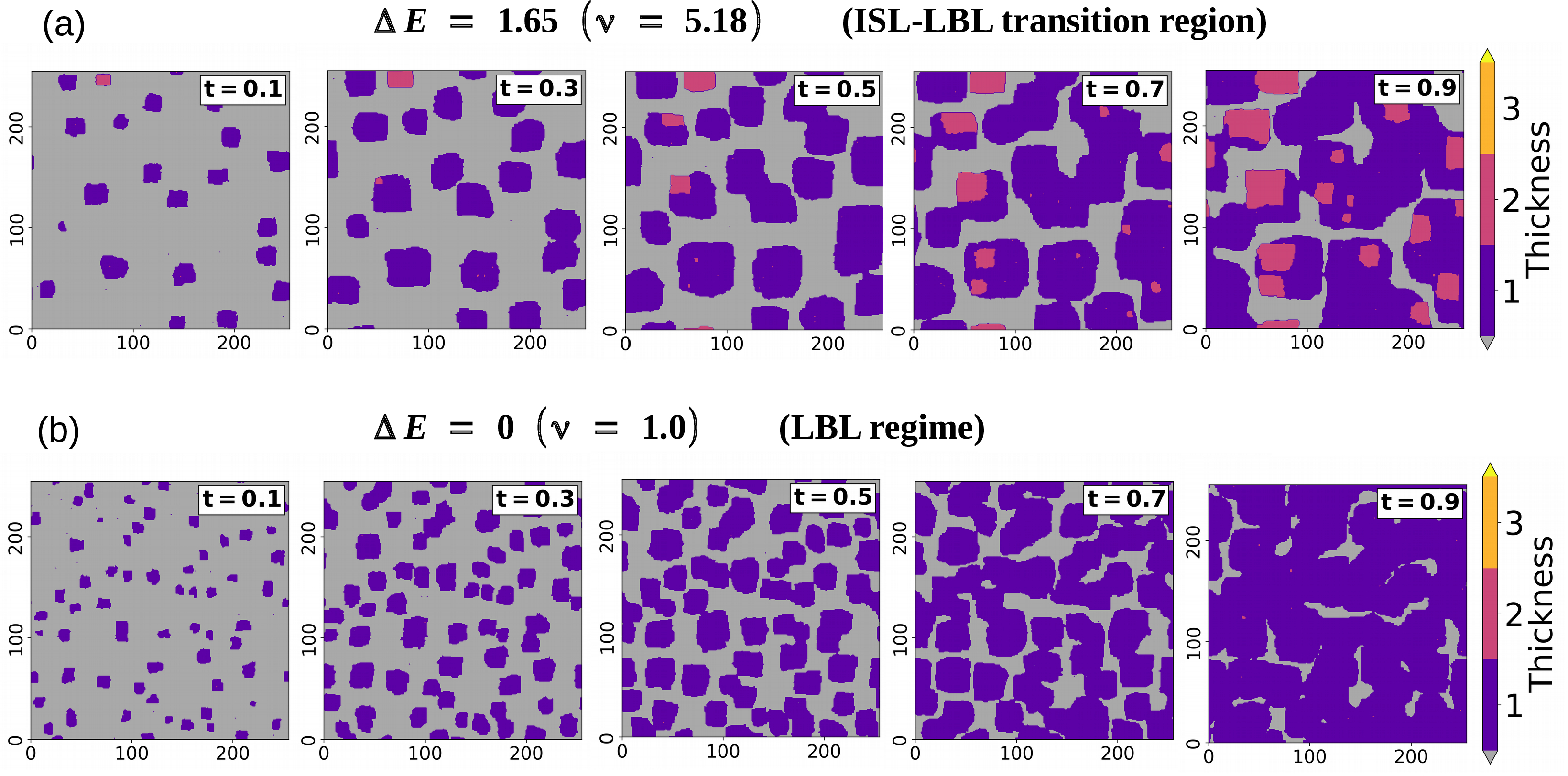}
  \caption{ Sequence of snapshots for $R=10^6$, $\Eb=5$. Upper panel: $\Delta E=1.65$ ($\nu=5.18$, midpoint of ISL--LBL transition region). Lower panel  $\Delta E=0$ ($\nus=1$), LBL regime.  }
  \label{fig:eb5_snaps2}
\end{figure*}

We illustrate this for the CV model with $R=10^6$ and $\Eb=5$.
Fitting the order parameter curve (see Fig.~\ref{fig:op_R6})  gives $\DEm \approx 1.69$ with a width $\approx 0.29$ which is already sizeable [Eq. (\ref{OPfit})].
The critical value of the ISL-LBL transition is thus $\Delta E_c \approx 2.27$.
We investigate the growth up to one ML for  4 parameters $\Delta E$ (marked in Fig.~\ref{fig:op_R6}).
The first ($\DE=2.71,\;\nu=15$) is in the ISL regime, the second ($\Delta E = 1.98,\;\nu=7.20$) and third  ($\Delta E = 1.65,\;\nu=5.18$) in the transition region between ISL and LBL, while the last  ($\Delta E = 0,\;\nu=1$) is homoepitaxial LBL growth.
Snapshots of deposits from $t=0.1$ to $0.9$ are shown in Figs.~\ref{fig:eb5_snaps1} and \ref{fig:eb5_snaps2}. 
For the first $\Delta E$ (Fig.~\ref{fig:eb5_snaps1}(a), deep in the ISL regime), 
at $t=0.1$ ML all islands are already converted to two--layer islands and the subsequent deposition only leads to lateral and vertical growth of these islands.
This corresponds to the picture of Sec.~\ref{sec:dynwet}.
For the second $\Delta E$ (Fig.~\ref{fig:eb5_snaps1}(b), already below the critical value), 
only two islands with two layers completed have formed in the early stages up to $t=0.1$.
A few more two--layer islands form until $t=0.3$ but already here and at later times one sees the nucleation of islands on the second layer which do not quickly cover the mother island beneath (as assumed by the scaling theory). Some of these nucleated islands need a time of several $0.1$~ML to cover the first layer island.
At $t\geq0.3$, there are actually not enough free monomers on the substrate in the capture zone around each first layer island for covering it with a second layer.
Therefore,  nucleation and growth of the island in the second layer must proceed with the help of directly deposited particles on the second layer. 

For the third value of $\Delta E$ (near $\DEm$, Fig.~\ref{fig:eb5_snaps2}(a)), 
one sees just one very quickly formed two--layer islands at $t=0.1$ and a few rather late nucleation events with slow growth from the secondary mechanism described above.
Finally, for the homoepitaxial growth with $\nu=1$ (LBL), no two-layer island appears.
In the scaling theory, the critical island size is infinity, and the secondary mechanism would set in very late, near $t=1$.
This would correspond to the traditionally discussed mechanism of second layer nucleation in epitaxial growth \citep{krugPRB2000}.

\end{appendix}

\newpage

\bibliography{interfaces}

\end{document}